%% file: tri4m.tex
\documentclass{article}
\usepackage{jheppub}
\usepackage{amsmath}
\usepackage{caption}
\usepackage{subcaption}
\usepackage{pict2e}
\usepackage{bbm}

\title{A non-planar two-loop three-point function beyond multiple polylogarithms}

\preprint{{MSUHEP-17-001, MITP/16-100, TTP17-001}}

\author[a,b]{Andreas von Manteuffel,}
\author[\,c]{Lorenzo Tancredi}

\affiliation[a]{Department of Physics and Astronomy, Michigan State University,
East Lansing, MI 48824, USA}
\affiliation[b]{PRISMA Cluster of Excellence, Johannes Gutenberg University,
55099 Mainz, Germany}
\affiliation[c] {Institute for Theoretical Particle Physics, KIT,
76128 Karlsruhe, Germany}

\emailAdd{manteuffel@pa.msu.edu}
\emailAdd{lorenzo.tancredi@kit.edu}

\keywords{Feynman integrals, Differential equations, Dispersion relations, Multiple Polylogarithms, Elliptic Integrals}

\abstract{
We consider the analytic calculation of a two-loop non-planar three-point function
which contributes to the two-loop amplitudes for $t \bar{t}$ production and 
$\gamma \gamma$ production in gluon fusion through a massive top-quark loop.
All subtopology integrals
can be written in terms of multiple polylogarithms over an irrational alphabet and we employ a new
method for the integration of the differential equations which does not rely on the rationalization
of the latter. 
The top topology integrals, instead, in spite of the absence of a massive
three-particle cut, cannot be evaluated
in terms of multiple polylogarithms and require the introduction of
integrals over complete elliptic integrals and polylogarithms. 
We provide one-fold integral representations for the solutions
and continue them analytically to all relevant  regions of the phase space in terms of real function, 
extracting all imaginary parts explicitly. The numerical evaluation of our expressions becomes  straightforward in this way.
}

\makeatletter
\renewcommand\@fpheader{}
\renewcommand\@journal{}
\makeatother

\newcommand{\e}{\epsilon}
\newcommand{\tm}{\tilde{m}}
\newcommand{\ud}{\mathrm{d}}
\DeclareMathOperator{\EK}{K}
\DeclareMathOperator{\EE}{E}
\DeclareMathOperator{\Li}{Li}
\DeclareMathOperator{\arcsec}{arcsec}
\DeclareMathOperator{\arccsc}{arccsc}
\DeclareMathOperator{\Cl}{Cl}

\newcommand{\R}{\mathcal{R}}
\newcommand{\I}{\mathcal{Q}}

\newcommand{\trianglecross}[3]{
\mbox{\parbox{3cm}{\hspace{-0.1cm}
\begin{picture}(2.5,1.4)
\thinlines
\put(0.5,0.7){\line(1,1){0.5}}
\put(0.5,0.7){\line(1,-1){0.5}}
\put(0.3,0.7){\vector(1,0){0.1}}
\put(1.8,0.2){\vector(1,0){0.1}}
\put(1.8,1.2){\vector(1,0){0.1}}
\put(1.5,1.2){\line(1,0){0.5}}
\put(1.5,0.2){\line(1,0){0.5}}
\put(0.25,0.9){\makebox(0,0)[b]{$#1$}}
\put(2.05,1.2){\makebox(0,0)[l]{$#2$}}
\put(2.05,0.2){\makebox(0,0)[l]{$#3$}}

\thicklines\linethickness{0.65mm}
\put(0,0.7){\line(1,0){0.5}}
\thicklines\linethickness{0.5mm}
\put(1,1.2){\line(1,0){0.5}}
\put(1,0.2){\line(1,0){0.5}}
\put(1.5,1.2){\line(-1,-2){0.5}}
\put(1.0,1.2){\line(1,-2){0.18}}
\put(1.5,0.2){\line(-1,2){0.18}}

\end{picture}
}}
\hfill}

\begin{document}
\maketitle
\setlength{\unitlength}{1.3cm} 
\allowdisplaybreaks

\section{Introduction}
The study of the mathematical properties of multiloop Feynman integrals
has received increasing attention in the last years both by
the physics and the mathematics communities. 
The high precision reached by the experimental measurements
carried out at the LHC, in fact, requires on the theory side the calculation of
multi- (typically two- or three-) loop corrections to various complicated $2 \to 1$,
$2 \to 2$ and $2 \to 3$ processes, including the exchange of
massless and massive virtual particles.
Indeed, unravelling their mathematical structures will be crucial
to handle the complexity of such calculations and, on the long run, it could help 
gain a deeper understanding of quantum field theory itself.

Impressive progress has been achieved in the last years in both directions.
The development of new techniques for the calculation
of multiloop Feynman integrals has rendered many previously out-of-reach
calculations feasible. Among these, a fundamental role has been played
by integration-by-parts reductions~\cite{Tkachov:1981wb,Chetyrkin:1981qh,Laporta:2001dd} 
and the differential equation method~\cite{Kotikov:1990kg,Remiddi:1997ny,Gehrmann:1999as},
augmented more recently by the use of a canonical basis~\cite{Kotikov:2010gf,Henn:2013pwa}.
\textsl{A posteriori}, a reason for the success of these techniques can be traced back to 
the almost concurrent ``discovery'' that the $\epsilon$ expansion of a rather large class of multiloop
(mainly \textsl{massless}) Feynman integrals can be expressed in terms of a very well understood
class of special functions, dubbed 
multiple polylogarithms~\cite{Goncharov,Goncharov:1998kja,Remiddi:1999ew,Gehrmann:2000zt}.
In particular the understanding of their algebraic and analytical properties, 
also thanks to the development
of the symbol and coproduct formalism~\cite{Goncharov:2005sla,Goncharov:2010jf,Brown:2011ik,Duhr:2011zq,Duhr:2012fh}, together
with stable routines for their evaluation on the whole complex 
plane~\cite{Gehrmann:2001pz,Gehrmann:2001jv,Vollinga:2004sn},
have played a crucial role in the last years.
Nevertheless, as it has been known for a long time, 
starting at the two-loop order the $\epsilon$ expansion of Feynman integrals can involve new mathematical structures 
which lie beyond the realm of multiple polylogarithms. The first and best studied example is that
of the two-loop massive Sunrise graph~\cite{Sabry,Broadhurst:1987ei,Bauberger:1994by,Bauberger:1994hx,Caffo:1998du,Laporta:2004rb,Bloch:2013tra,Remiddi:2013joa,
Adams:2013nia,Adams:2014vja,Adams:2015gva,Adams:2015ydq,Bloch:2016izu,Remiddi:2016gno}.
Its imaginary part in $d=4$ space-time dimensions is known to be expressible as linear combination of
complete elliptic integrals of first and second kind. The real part of the graph can be reconstructed by a dispersive relation and involves integrals over elliptic integrals.
In the last years quite some effort has been spent on studying the new functions
appearing in the Sunrise graph, trying to extend the properties
of multiple polylogarithms to embrace also elliptic generalizations of the latter~\cite{Bloch:2013tra,Adams:2013nia,Adams:2014vja,Adams:2015gva,Adams:2015ydq,Bloch:2016izu}. 
Much progress has been made but a complete understanding of these functions is still missing,
in particular regarding their analytic continuation and numerical evaluation.
Moreover, it is not clear whether they can be easily used to express also graphs with
more complicated kinematics (i.e.\ three- and four-point functions).
Last but not least, the interplay of these functions with the simpler multiple polylogarithms
and their iterative structure remains unclear in context with the method of differential equations.
There, simpler integrals appear  as inhomogeneous
terms in the differential equations of more complicated integrals. The solution of these equations
requires on the one hand the solution of the homogeneous equations and, on the other, the integration over
the inhomogeneous piece. Multiple polylogarithms 
(and in general Chen iterated integrals~\cite{Chen:1977oja})
are particularly well suited for this scope, as they can be defined as iterative integrals over a set of
differential forms.
 Of course, it is a priori unclear whether such a feature can be expected to 
hold for arbitrarily complicated Feynman graphs.

More recently, a new picture has started to emerged, where a generalization of this approach 
to more complicated cases becomes possible. 
When dealing with a system of coupled differential equations, the most non-trivial
step consists in solving the homogeneous part of the system. In~\cite{Primo:2016ebd} it was shown
that, for arbitrarily complicated cases, integral representations of the homogeneous
solutions can be found by computing the maximal cut of the graph under consideration.
Once the homogeneous solution is known, an integral representation for the
inhomogeneous solution is provided by Euler's variation of constants. Even if usually such integrals 
cannot be expressed in terms of known special functions,
from a practical point of view we are particularly interested in obtaining results that allow
for fast and reliable numerical evaluations over the physical phase space.
In this context, it was shown in~\cite{Remiddi:2016gno}
that the study of the imaginary part and of the corresponding
 dispersion relations of Feynman integrals within the differential equations framework
 can facilitate obtaining compact one-fold
integral representations for the two-loop massive Sunrise and the Kite integral\footnote{Note that more recently it was shown that the Kite integral can also be re-written in terms of elliptic polylogarithms~\cite{Adams:2016xah}.}.
There are indications that more complicated Feynman integrals with different and unrelated kinematics 
can be casted in a similar form, see for example~\cite{Aglietti:2007as,CaronHuot:2012ab} and more recently a set of two-loop planar double boxes relevant for $H+$jet production~\cite{Bonciani:2016qxi}.

The new ideas summarized above provide us with the tools required to start successfully 
studying more examples
of relevant Feynman diagrams whose $\epsilon$ expansions do not evaluate to multiple polylogarithms.
In this paper, we focus on a three-point non-planar topology 
which is relevant for the two-loop QCD
 corrections to the processes $gg\to t\bar{t}$
 and $g g \to \gamma \gamma$ 
 through a massive top loop. 
Similar integrals
appear in the non-planar sectors of $H+$jet and $HH$ production. 
The rest of the paper is organized as follows.
In Section~\ref{sec:notation} we describe the problem and establish the notation.
We continue in Section~\ref{sec:subtopos} showing how the $\epsilon$ expansion
of the simpler subtopologies can be integrated in terms of multiple polylogarithm
with a new algorithm based on the differential equations in canonical form.
In Section~\ref{sec:toptopo} we consider the top topology, which cannot be integrated in terms 
of multiple polylogarithms. We show how to solve the coupled system of
differential equations satisfied by its two master integrals based on
information extracted from the maximal cut;
we then compare it to the result obtained by solving directly the second order differential equation.
We thoroughly study the required functions in Section~\ref{sec:toptopo} 
and Appendix~\ref{sec:ancont}, where
we also show how to analytically continue our solution to all relevant regions of the phase space.
Finally in Section~\ref{sec:inhomsol} we use the results of the previous sections
to build up the inhomogeneous solution
for the top topology in analytic form. We write the results 
as one-fold integrals over rational and irrational functions, 
multiple polylogarithms and complete elliptic integrals of first and second kind.
We then draw our conclusions in Section~\ref{sec:conclusions}.

\section{Differential equations}\label{sec:notation}
We consider the Feynman integrals family 
$I_{a_1,a_2,a_3,a_4,a_5,a_6,a_7}$ defined as
\begin{align}
& \trianglecross{q}{p_1}{p_2} = I_{a_1,a_2,a_3,a_4,a_5,a_6,a_7} \Big|_{a_7 < 0} \nonumber \\
& = \int \, 
\frac{ \ud^dk\, \ud^d l \;\; (m^2)^{2\epsilon} N_\epsilon^2 \;\; (k^2)^{-a_7}}{[(k-p_1)^2]^{a_1} [(l-p_1)^2-m^2]^{a_2} [(k+p_2)^2]^{a_3}
[(k-l+p_2)-m^2]^{a_4} [(k-l)-m^2]^{a_5}[(l^2-m^2)^{a_6}]}\,, \label{eq:topo}
\end{align}
where
\begin{equation}
N_\epsilon = \frac{1}{i \pi^{2-\epsilon} \Gamma(1+\epsilon)}
\end{equation}
and $\epsilon=(4-d)/2$.
We stress that the triangle topology above is obtained by limiting ourselves to negative
powers of the last propagator, i.e.\ $a_7<0$.
Four of the six propagators are massive (thick lines), while only one of the three external legs is
off-shell, i.e.\ $p_1^2 = p_2^2 = 0$ and $q^2 =(p_1+p_2)^2 = s$\,.
Integrals in family~\eqref{eq:topo} with different values of the exponents $a_i$
may be reduced to master integrals using integration by parts reductions
as implemented for example in Reduze\;2~\cite{vonManteuffel:2012np,Studerus:2009ye,Bauer:2000cp,fermat}.
As a result we find 11 different master integrals: 9 for the subtopologies
and 2 for the top topology.

As we will see explicitly, the nine master integrals of the subtopologies
can be expressed in terms of multiple polylogarithms only.
For them we choose a canonical basis
using the method described in~\cite{Gehrmann:2014bfa}.\footnote{For recent related work on the analysis of master
integrals and the construction of a canonical basis see
\cite{Lee:2013hzt,Argeri:2014qva,Lee:2014ioa,Tancredi:2015pta,Ita:2015tya,
Meyer:2016slj,Georgoudis:2016wff,Prausa:2017ltv,Gituliar:2017vzm}.
}
The remaining two integrals, instead, contain elliptic generalizations of the multiple polylogarithms and, therefore, it is not possible to find for them a canonical basis as defined in~\cite{Henn:2013pwa}.\footnote{We stress here that 
a canonical basis as defined in~\cite{Henn:2013pwa} requires not only a specific factorization of
the $\epsilon$ dependence from the kinematics, but also, as an additional condition on the matrix of the system,
a d-$\log$ form for the differential equations.
In the general case the first condition could still be realized, while the d-$\log$ form
condition might have to be generalized to differentials of more complicated functions.}
This can be motivated using the ideas presented in~\cite{Tancredi:2015pta}. There it was shown
for various examples that the possibility of decoupling the differential equations in the limit $d \to 4$ (and therefore of
writing the result in terms of multiple polylogarithms) is signaled by a degeneracy
of the integration by parts identities in the limit of even numbers of dimensions, 
$d \to 2\,n$ with $n \in \mathbb{N}$.

We first consider the part of the Euclidean region where $s<-4\,m^2$ and employ the
following basis
\begin{align}
m_{1} &= \epsilon^2\, I^{2,10}_{0,2,0,2,0,0,0}\,, &
m_{2} &= \epsilon^2\, s\, I^{3,7}_{2,2,1,0,0,0,0} \nonumber \\
m_{3} &= \epsilon^2\,s\, I^{3,22}_{0,2,1,0,2,0,0}\,, & 
m_{4} &= \epsilon^2\, \sqrt{s(s-4\,m^2)}\, \left[ I^{3,22}_{0,2,2,0,1,0,0} 
            + \frac{1}{2}  I^{3,22}_{0,2,1,0,2,0,0} \right] \nonumber \\
m_{5} &= \epsilon^3\, s\, I^{4,15}_{1,2,1,1,0,0,0}\,, &
m_{6} &= \epsilon^2\,\sqrt{s(s+4\,m^2)} \left[ s \, I^{4,15}_{2,2,1,1,0,0,0} 
            - \frac{\epsilon}{2\,m^2 (1+2 \epsilon)}\, I^{2,10}_{0,2,0,2,0,0,0}\right] \nonumber \\
m_{7} &= \epsilon^3\, s\, I^{4,30}_{0,2,1,1,1,0,0}\,, &
m_{8} &= \epsilon^4\, s\, I^{5,59}_{1,1,0,1,1,1,0}\,, \qquad\;
m_{9} = \epsilon^4 \, s\, I^{5,31}_{1,1,1,1,1,0,0}\,, \nonumber \\
m_{10} &= \e^4 s^2 I^{6,63}_{1, 1, 1, 1, 1, 1, 0}\,, &
m_{11} &= \e^4 \frac{s^2 (s + 16 m^2)}{2 (1 + 2 \e)} I^{6,63}_{1, 2, 1, 1, 1, 1, 0}\,,
\label{eq:basis}
\end{align}
where we added the number of different denominators and {Reduze}'s sector id as a superscript to
the $I$ integrals for easier identification.

In order to write down the system of differential equations we introduce the
massless ratio
\[
x = -s/m^2\,.
\]
We introduce the vector $\vec{m} = (m_1, \ldots, m_9)$ for the master
integrals of the subtopologies, which fulfil canonical differential
equations. In matrix notation we obtain
\begin{equation} \label{eq:deqcan}
\ud \vec{m} = \epsilon\, \sum_{i=1}^5 \ud\,\ln(l_i(x))\,A_i \,\vec{m}
\end{equation}
with the letters
\begin{equation}\label{eq:alphabet}
 l_1 = \sqrt{x},\quad
 l_2 = \tfrac{1}{2}(\sqrt{x} + \sqrt{x+4}),\quad
 l_3 = \sqrt{x + 4},\quad
 l_4 = \tfrac{1}{2}(\sqrt{x} + \sqrt{x - 4}),\quad
 l_5 = \sqrt{x - 4}
\end{equation}
and the matrices
\begin{alignat}{3}
A_1 &= \left( \begin{smallmatrix}
0& 0& 0& 0& 0& 0& 0              & 0& 0\\ 
0&-2& 0& 0& 0& 0& 0              & 0& 0\\
0& 0& 2& 0& 0& 0& 0              & 0& 0\\
0& 0& 0&-2& 0& 0& 0              & 0& 0\\
0& 0& 0& 0& 2& 0& 0              & 0& 0\\
0& 0& 0& 0& 0&-2& 0              & 0& 0\\
0& 0& 2& 0& 0& 0& 0              & 0& 0\\
0& 0& 0& 0& 0& 0& 2              & 0& 0\\
0& 0& 0& 0& 2& 0&-1& 0& -2
\end{smallmatrix} \right), &\quad
A_2 &= \left( \begin{smallmatrix}
0& 0& 0& 0& 0& 0& 0& 0& 0\\ 
0& 0& 0& 0& 0& 0& 0& 0& 0\\
0& 0& 0& 4& 0& 0& 0& 0& 0\\
2& 0& -6& 0& 0& 0& 0& 0& 0\\
0& 0& 0& 0& 0& 0& 0& 0& 0\\
0& 0& 0& 0& 0& 0& 0& 0& 0\\
0& 0& 0& 0& 0& 0& 0& 0& 0\\
0& 0& 0& 0& 0& 0& 0& 0& 0\\
0& 0& 0& 0& 0& 0& 0& 0& 0
\end{smallmatrix} \right), &\quad
A_3 &= \left( \begin{smallmatrix} 
0& 0& 0& 0& 0& 0& 0& 0& 0\\ 
0& 0& 0& 0& 0& 0& 0& 0& 0\\
0& 0& 0& 0& 0& 0& 0& 0& 0\\
0& 0& 0&-6& 0& 0& 0& 0& 0\\
0& 0& 0& 0& 0& 0& 0& 0& 0\\
0& 0& 0& 0& 0& 0& 0& 0& 0\\
0& 0& 0& 0& 0& 0& 0& 0& 0\\
0& 0& 0& 0& 0& 0& 0& 0& 0\\
0& 0& 0& 0& 0& 0& 0& 0& 0
\end{smallmatrix} \right),
\nonumber\\
A_4 &= \left( \begin{smallmatrix} 
0& 0& 0& 0& 0& 0& 0& 0& 0\\ 
0& 0& 0& 0& 0& 0& 0& 0& 0\\
0& 0& 0& 0& 0& 0& 0& 0& 0\\
0& 0& 0& 0& 0& 0& 0& 0& 0\\
0& 0& 0& 0& 0& 2& 0& 0& 0\\
2& 2& 0& 0&-6& 0& 0& 0& 0\\
0& 0& 0& 0& 0& 0& 0& 0& 0\\
0& 0& 0& 0& 0& 0& 0& 0& 0\\
0& 0& 0& 0& 0& 0& 0& 0& 0
\end{smallmatrix} \right), &\quad
A_5 &= \left( \begin{smallmatrix} 
0& 0& 0& 0& 0& 0& 0& 0& 0\\ 
0& 0& 0& 0& 0& 0& 0& 0& 0\\
0& 0& 0& 0& 0& 0& 0& 0& 0\\
0& 0& 0& 0& 0& 0& 0& 0& 0\\
0& 0& 0& 0& 0& 0& 0& 0& 0\\
0& 0& 0& 0& 0&-4& 0& 0& 0\\
0& 0& 0& 0& 0& 0& 0& 0& 0\\
0& 0& 0& 0& 0& 0& 0& 0& 0\\
0& 0& 0& 0& 0& 0& 0& 0& 0
\end{smallmatrix} \right).
&&
\end{alignat}
The root appearing in the normalization of $m_4$ leads to $l_2$ and $l_3$,
while the root appearing in the normalization of $m_6$ leads to $l_4$ and $l_5$.
We note that, as long as we limit ourselves to $m_1,\ldots,m_9$,
these pairs of roots never mix and we could rationalize them separately
with two different changes of variables. 
The two structures, nevertheless, mix up once considering the differential equations for the
two master integrals of the top sector, $m_{10}$ and $m_{11}$.

With the basis given in~\eqref{eq:basis}, the differential equations for the master integrals
of the top topology read
\begin{align}
\frac{d}{dx} \left(\begin{matrix} m_{10} \\ m_{11} \end{matrix} \right) =
B(x) \left(\begin{matrix} m_{10} \\ m_{11} \end{matrix} \right) 
+ \epsilon\, D(x) \left(\begin{matrix} m_{10} \\ m_{11} \end{matrix} \right) 
+ \left(\begin{matrix} N_{10}(\epsilon;x) \\ N_{11}(\epsilon;x) \end{matrix} \right) \label{eq:topsys}
\end{align}
where $B(x)$ and $D(x)$ are two $2 \times 2$ matrices that do not depend on $\epsilon$,
while $N_{10}(\epsilon;x)$ and $N_{11}(\epsilon;x)$  contain the dependence on the
simpler subtopologies.
The matrices of the homogeneous part are
\begin{align}
B(x) = \left( \begin{array}{ccc} 0 & & \frac{1}{2(x-16)} - \frac{1}{2\,x} \\&\\
                                        \frac{1}{2\,x} & & \frac{1}{x}  \end{array}\right)\,, 
                                        \qquad 
D(x) = \left( \begin{array}{ccc} - \frac{2}{x} & & \frac{1}{x-16} - \frac{1}{x} \\&\\
                                        \frac{2}{x} & & \frac{1}{x} - \frac{1}{x-16} \end{array}\right)\,,
\end{align}
while the non-homogeneous terms read
\begin{align}
&N_{10}(\e; x) = 0\,, \nonumber \\
&N_{11}(\e; x) = \frac{\e^2}{1+2\e} \left( \frac{5}{4} \,m_3 - \frac{7}{2} \,m_5 
- \frac{3}{2} \frac{x}{\sqrt{x(x-4)}} \, m_6 + 2\, m_7 + m_9\, \right)\,. \label{eq:nonhom}
\end{align}
From the latter it is clear that in the top topology all letters
appear simultaneously.
In our normalization the integrals $m_j = m_j(\e; x)$ $(j=1,\ldots,11)$ have a
Taylor series expansion at $\e=0$,
\begin{equation}\label{mepsexpansion}
m_j(\e; x) = \sum_{n=0}^{\infty} m_j^{(n)}(x)\; \epsilon^n\qquad\text{for}~j=1,\ldots,11,
\end{equation}
and we wish to calculate the expansion coefficients up to $n=4$.

\section{Integration of the subtopologies}\label{sec:subtopos}

We first consider the evaluation of the subtopologies $m_1,\ldots,m_9$.
Inserting the expansion~\eqref{mepsexpansion} into the differential
equation~\eqref{eq:deqcan} and comparing coefficients of powers in $\e$ gives
\begin{equation}\label{deqexpansion}
\ud m_j^{(n)}(x) = \sum_{i=1}^5 \ud\,\ln(l_i(x))\,A_i \,m_j^{(n-1)}(x)\qquad\text{for~}n=0,\ldots,\infty,
\end{equation}
which allows us to solve fully decoupled systems of differential equations
order by order $n$ bottom-up.
The differential equations~\eqref{deqexpansion} have the form of the
differential equations of multiple polylogarithms with root-valued
letters.
We find it useful to solve the full vector $\vec{m}$ in a uniform setup,
where we consider all letters at the same time.
This prevents a rationalization of the $x$ dependence and
therefore a traditional integration of the differential equations.

Instead, we construct an ansatz for the solution using suitable target
functions and require it to fulfil the differential equation~\eqref{deqexpansion}.
In other words, we \emph{integrate the symbol} of the loop integrals.
The basis of our function construction is the Duhr-Gangl-Rhodes algorithm~\cite{Duhr:2011zq},
which we extend here to the case of an irrational alphabet.
For simpler cases it is typically not strictly necessary to work with
a canonical basis in order to be able to integrate the differential equations,
provided they (partially) decouple.
In contrast to this, our strategy here crucially relies on the
differential equation~\eqref{deqexpansion} being in canonical form,
such that we can read off the symbol letters of the solution. A similar approach
for the integration of the differential equations has been used in~\cite{Bonciani:2016qxi}
to calculate solutions through to weight two.

To construct our solutions in the region $s<-4m^2$ through to weight four
we choose the functions
\begin{equation}
\ln,\quad \Li_2,\quad \Li_3,\quad \Li_4,\quad \Li_{2,2}
\end{equation}
with suitable arguments.
We look for functions which do not introduce symbol letters beyond the alphabet
\eqref{eq:alphabet} of the differential equation~\eqref{deqexpansion}.
For the arguments of $\ln$ we can just choose the original letters themselves.

A classical polylogarithm $\Li_j(z)$ has symbol entries $z$ and $1-z$.
Choosing $z$ to be a power product of our original letters,
\begin{equation}
\label{powerproduct}
c^{a_0} l_1^{a_1} \cdots l_5^{a_5},
\end{equation}
with $c$ a rational number, ensures $z$ to factorize over our alphabet.
Here, it is enough to consider $a_n\in \mathbbm{Z}$ and $c^{a_0}=\pm 1$.
We stress that this is too constraining in general.
In particular, depending on the (choice of) letters it may be necessary
to also consider roots of letters for factorization and allow the $a_n$ to be
non-integer rational numbers.
Note that there is an ambiguity related to the freedom of redefining the 
letters by expressions
of the form \eqref{powerproduct}.
In our case, the definition of our letters absorbs any further explicit
roots.

In addition to the argument $z$ of a polylogarithm we also require
$1-z$ to factorize in the sense \eqref{powerproduct} over our original alphabet.
This condition is challenging to check with computer algebra systems, in particular
for multivariate applications.
We therefore resort to decompositions found by inserting numbers
for the variables and applying heuristic integer relation searches to the logarithms
of the involved factors~\cite{weihsprivcomm,Lenstra1982,PARI2} (more details will be given in~\cite{rmsavm}).
Our heuristic search produces this admissible set:
\begin{align}
\{&1/(l_1 l_4),\; l_4/l_1,\; 1/(l_2 l_3),\; l_3/
 l_2,\; -(1/(l_1 l_2)),\; -(1/(l_4 l_5)),\; l_5/l_4,\; 1/l_4^2,\; l_4/
 l_5,\; l_2/l_3,\; l_2/l_1,\; l_4^2,\; 1/
 l_2^2,\;  \nonumber\\
 &-(1/l_4^2),\; 
-l_4 l_5,\; -(1/l_2^2),\; l_2 l_3,\; l_1/l_4,\; 
 l_1 l_4,\; -l_4^2,\; l_1/l_2,\; 
 l_2^2,\; -l_2^2,\; -l_1 l_2,\; -(1/(l_1 l_2^2 l_3)),\; -(1/(
  l_1 l_4^2 l_5)),\;\nonumber\\
 & l_4^2/(l_1 l_5),\; 1/l_2^4,\;
 l_2^2/(
 l_1 l_3),\; 1/l_4^4,\; l_4^4,\; (l_1 l_5)/l_4^2,\; -l_1 l_4^2 l_5,\; 
 l_2^4,\; (l_1 l_3)/l_2^2,\; -l_1 l_2^2 l_3\},
\end{align}
which is closed under $z \to 1-z$ and $z \to 1/z$, where $z$ is an element of the set.
For $\Li_{2,2}(z_1,z_2)$ we need to make sure that the symbol factors
$z_1$, $1-z_1$, $z_2$, $1-z_2$ and $z_1-z_2$ do not introduce new letters.
Our construction proceeds along the lines described above, but we 
do not list the generic result for the argument set here since it is too lengthy.

We concentrate here on the part of the Euclidean region where $s<-4m^2$
and require all functions to be real valued for $x>4$.
Indeed, we find that all of these constraints can be satisfied and a solution
in terms of such functions can be found, which satisfies the differential
equations.
Our functions read
\begin{align}
\{
 &\ln(l_1), \ln(l_2), \ln(l_3), \ln(l_4), \ln(l_5),
 \Li_2(l_2^{-2}), \Li_2(1/(l_2 l_3))), \Li_2(-l_4^{-2}), \Li_2(-1/(l_4 l_5)),
 \Li_3(l_2^{-2}), 
 \nonumber \\
 &\Li_3(l_1/l_2), \Li_3(-1/(l_1 l_2^2 l_3)), 
 \Li_3(1/(l_2 l_3))), \Li_3(l_2/l_3)), \Li_3(-l_4^{-2}), \Li_3(l_4^{-2}),
 \Li_3(1/(l_1 l_4)),
 \nonumber \\
 & \Li_3(-1/(l_4 l_5))), \Li_3(l_1 l_5)/l_4^2), 
 \Li_4(-l_2^{-2}), \Li_4(l_2^{-2}), \Li_4(-1/(l_1 l_2)), \Li_4(l_1/l_2), 
 \Li_4(-1/(l_1 l_2^2 l_3)),
 \nonumber \\
 &
 \Li_4(1/(l_2 l_3)), \Li_4(l_2/l_3), \Li_4(l_1 l_3)/l_2^2), \Li_4(-l_4^{-2}), \Li_4(l_4^{-2}), 
 \Li_4(1/(l_1 l_4)), \Li_4(l_4/l_1), \Li_4(-1/(l_1 l_4^2 l_5)), 
 \nonumber \\
 &
 \Li_4(-1/(l_4 l_5)), \Li_4(l_1 l_5/l_4^2), \Li_4(l_5/l_4), \Li_{2,2}(-1, -l_2^{-2}),
 \Li_{2,2}(-1, -l_4^{-2}), \Li_{2,2}(-1/(l_1 l_2), l_1/l_2),
 \nonumber \\
 & 
 \Li_{2,2}(1/(l_1 l_4), l_1/l_4) 
\}\,.
\end{align}
In order to fix the integration constants, we use the massive tadpole
and the massless bubbles as input as well as various regularity conditions.

We find very compact results for the final solutions of the subtopology
integrals in terms of our choice of functions.
These results will be a necessary building block to construct the full solution of the top topology.
The explicit form of the results can be found in Appendix~\ref{app:subtopos}
and, in electronic format, in the ancillary file of the arXiv submission of this paper.

\section{Integration of the top topology}\label{sec:toptopo}

\subsection{Homogenous solution}

Having the expressions for all subtopologies in terms of multiple polylogarithms,
we can now proceed and study the differential equations~\eqref{eq:topsys} for the top sector.
We start by expanding in $\e$ the inhomogeneous terms~\eqref{eq:nonhom}, up to 
$\e^4$
\begin{equation}
N_{j}(\epsilon;x) = \sum_{a=0}^4 \, N_j^{(a)}(x)\,\epsilon^a\,,\qquad \mbox{for}\;j=10,11\,.
\end{equation}
Substituting explicitly the subtopologies we find the remarkably simple expressions
\begin{align}
&N_{10}^{(0)}(x) = 0\,, \quad N_{10}^{(1)}(x) = 0\,, \quad N_{10}^{(2)}(x) = 0\,,\quad N_{10}^{(3)}(x) =0\, \quad
N_{10}^{(4)}(x) = 0\,,\nonumber \\
&N_{11}^{(0)}(x) = 0\,, \quad N_{11}^{(1)}(x) = 0\,, \quad N_{11}^{(2)}(x) = 0\,,\quad N_{11}^{(3)}(x) =0 \,,\nonumber \\
&N_{11}^{(4)}(x) = 5 \ln^2 (l_2)
- l_1\; \frac{3/2\, \zeta_2   
+ 3 \ln^2( l_4) + 3 \Li_2( -1/l_4^2 )}{ l_5}\,, \label{eq:nonhomexp}
\end{align}
where the letters $l_j$ have been defined above~\eqref{eq:alphabet}.
We can now expand the right- and left-hand-side of~\eqref{eq:topsys} and we find, 
at any order $n$, 
\begin{align}
\frac{d}{dx} \left(\begin{matrix} m^{(n)}_{10} \\ m_{11}^{(n)} \end{matrix} \right) =
B(x) \left(\begin{matrix} m^{(n)}_{10} \\ m^{(n)}_{11} \end{matrix} \right) 
+  D(x) \left(\begin{matrix} m^{(n-1)}_{10} \\ m^{(n-1)}_{11} \end{matrix} \right) 
+ \left(\begin{matrix} N^{(n)}_{10}(x) \\ N^{(n)}_{11}(x) \end{matrix} \right)\,.\label{eq:topsysex}
\end{align}
Comparing with~\eqref{eq:nonhomexp}, we see that we get a non trivial
non-homogeneous term only at order $\e^4$, which is also the maximum
order in the expansion we are interested in.
In fact, the integrals $I^{6,63}_{1, 1, 1, 1, 1, 1, 0}$
and $I^{6,63}_{1, 2, 1, 1, 1, 1, 0}$ are finite, which can be checked
e.g.\ with the methods of \cite{Panzer:2014gra,vonManteuffel:2014qoa} implemented in {Reduze\;2.1},
such that the first non-vanishing coefficients of $m_{10}$ and $m_{11}$ occur at order $\e^4$.

The first step to solve~\eqref{eq:topsysex} is to find a solution of its homogeneous part,
which at every order in $\epsilon$ is a coupled system of two first-order linear differential equations
\begin{equation}
\frac{d}{dx} \left(\begin{matrix} M_{10} \\ M_{11} \end{matrix} \right) =
B(x) \left(\begin{matrix} M_{10} \\ M_{11} \end{matrix} \right)\,, \label{eq:tophom}
\end{equation}
where we use a capital letter $M$ instead of $m$ to describe the homogeneous part
of the solution and drop the superscript $(n)$, since the form of the equation does not depend on $n$.
In order to solve the system we need to find two independent set of solutions, i.e.\ a $2 \times 2$ matrix
\begin{align}
G(x) &= \left(\begin{matrix} I_1(x) & J_1(x)  \\ I_2(x) & J_2(x) \end{matrix}\right) \label{eq:matsol}
\end{align}
such that
\begin{equation}
\frac{d}{dx} G(x) = B(x) \, G(x)\,. \label{eq:deqmat}
\end{equation}
The inverse of the matrix $G(x)$ reads
\begin{equation}
G^{(-1)}(x) = \frac{1}{W(x)} \left( \begin{array}{cc} J_2(x) & -J_1(x) \\ -I_2(x) & I_1(x)  \end{array}\right)\,, \label{eq:invG}
\end{equation}
where $W(x)$ is the Wronskian of the solutions, $W(x) = I_1(x)J_2(x) - J_1(x)I_2(x)$\,.
If the matrix $G(x)$ is known, one can perform at every order in $\epsilon$ the rotation
\begin{align}
\left(\begin{matrix} m^{(n)}_{10} \\ m^{(n)}_{11} \end{matrix} \right)
&= G \left(\begin{matrix} \tilde{m}^{(n)}_{10} \\ \tilde{m}^{(n)}_{11} \end{matrix} \right) \label{eq:rotation}
\end{align}
such that the new functions $\tilde{m}^{(n)}_{10}, \tilde{m}^{(n)}_{11}$ fulfil the differential equations
\begin{align}
\frac{d}{dx} \left(\begin{matrix} \tilde{m}^{(n)}_{10} \\ \tilde{m}_{11}^{(n)} \end{matrix} \right) =
G^{(-1)}(x)\,D(x)\, G(x) \left(\begin{matrix} \tilde{m}^{(n-1)}_{10} \\ \tilde{m}^{(n-1)}_{11} \end{matrix} \right) 
+ G^{(-1)}(x) \left(\begin{matrix} N^{(n)}_{10}(x) \\ N^{(n)}_{11}(x) \end{matrix} \right)\, \label{eq:toprot}
\end{align}
whose solution can be now written by quadrature as
\begin{align}
\left(\begin{matrix} \tilde{m}^{(n)}_{10}(y) \\ \tilde{m}_{11}^{(n)}(y) \end{matrix} \right) =
\int_{y_0}^y \, dx\, G^{(-1)}(x)\,D(x)\, G(x) \left(\begin{matrix} \tilde{m}^{(n-1)}_{10} \\ \tilde{m}^{(n-1)}_{11} \end{matrix} \right) 
+ \int_{y_0}^y \, dx\, G^{(-1)}(x) \left(\begin{matrix} N^{(n)}_{10}(x) \\ N^{(n)}_{11}(x) \end{matrix} \right)
+ \left(\begin{matrix} c^{(n)}_{10} \\ c^{(n)}_{11} \end{matrix} \right) \label{eq:soltoprot}
\end{align}
where $c_{10}^{(n)}$ and $c_{11}^{(n)}$ are two integration constants and $y_0$ is a suitable boundary for the integration.

In order to find an explicit solution for the matrix $G(x)$ it is useful to recast~\eqref{eq:tophom} as a 
second order differential equation for the first master integral
\begin{equation}
\frac{d^2 \, M_{10}(x) }{dx^2} + \left( \frac{1}{x-16}\right)\frac{d \, M_{10}(x) }{dx} 
+ \frac{1}{64}\left( \frac{1}{x}  + \frac{16}{x^2} - \frac{1}{x-16} \right)M_{10}(x) = 0\,. \label{eq:secdeqm}
\end{equation}
A general way to solve this equation is to compute the maximal cut of $M_{10}(x)$
as explained in detail in~\cite{Primo:2016ebd}. In that reference it was shown that
the maximal cut of $M_{10}(x)$ computed on a specific integration contour can be written 
(up to an irrelevant overall numerical constant) as
\begin{align}
{\rm Cut_1}(M_{10}(x)) = \sqrt{ \frac{x}{x-16} } \EK\left( \frac{x}{x-16}\right)\,, \label{eq:cut1}
\end{align}
where we introduced the complete elliptic integral of the first kind defined as
\begin{equation}
\EK(z) = \int_0^1 \frac{dx}{\sqrt{(1-x^2)(1-\,z\,x^2)}} \,, \quad 
\mbox{for} \quad z \in \mathbb{C} \quad \mbox{and}\quad \Re{(z)} < 1\,. \label{eq:ellint}
\end{equation}
It is straightforward to verify that~\eqref{eq:cut1} satisfies~\eqref{eq:secdeqm}.
A second independent solution can be obtained integrating on an independent contour.
As explained in~\cite{Primo:2016ebd}, this can be avoided in the case of elliptic integrals,
as the second solution is simply given by
\begin{align}
{\rm Cut_2}(M_{10}(x)) = \sqrt{ \frac{x}{x-16} } \EK\left( 1- \frac{x}{x-16}\right)\,. \label{eq:cut2}
\end{align}
The two solutions~\eqref{eq:cut1} and \eqref{eq:cut2} completely determine the first row of the matrix~\eqref{eq:matsol}.

Before proceeding, it is instructive to try to solve Eq.~\eqref{eq:secdeqm} directly, since in this case the equation is particularly
simple. We define the auxiliary function
\begin{equation}
f(x) = \frac{1}{\sqrt{x}}\, M_{10}(x)\,,
\end{equation}
which fulfills the differential equation
\begin{equation}
\frac{d^2 \, f(x) }{dx^2} + \left( \frac{1}{x} + \frac{1}{x-16} \right) \frac{d \,f (x) }{dx} 
+ \frac{1}{64} \left(\frac{1}{x-16} - \frac{1}{x} \right)\, f(x) = 0 \,. \label{eq:secdeq}
\end{equation}
Equation~\eqref{eq:secdeq} is now in standard form and its solution can be 
easily written in terms
of the complete elliptic integral of the first kind~\eqref{eq:ellint}.
Since the differential equation~\eqref{eq:secdeq} has regular singular points in $x=0$, $x=16$
and $x=\pm \infty$, we must consider the solution of the latter in the three intervals $0<x<16$, 
$16<x<\infty$ and $-\infty<x<0$\,.
We start by considering the region $0<x<16 $\, i.e.\ $-16\,m^2 < s < 0$.
This region is non physical and the master integrals $m_{10}$, $m_{11}$ are real there.
Following the same procedure, we can build up similar solutions in the remaining
regions, i.e.\ the part of the Euclidean region where  $-\infty < s < -16\,m^2$ and
the physical region $s>0$.
We derive explicitly those solutions in Appendix~\ref{sec:ancont}.
We use the notation $I_j^{(a,b)}(x)$ and $J_j^{(a,b)}(x)$, with $j=1,2$, for the two pairs
of solutions valid in the region $a<x<b$, and consequently $G^{(a,b)}(x)$ for the matrix of the solutions
in the same region.
The general solution of the homogeneous second order differential equation for $0<x<16$
can be written as
\begin{equation}
\label{eq:fsol}
f(x) = c_{1}\, \EK \left(  \frac{x}{16} \right) + c_{2} \EK \left( 1 - \frac{x}{16} \right)\,,
\end{equation}
which determines the first row of the matrix~\eqref{eq:matsol} as
\begin{align}
I_1^{(0,16)}(x) = \sqrt{x}\, \EK \left(  \frac{x}{16} \right)\,, \qquad J_1^{(0,16)}(x) = 
\sqrt{x}\, \EK \left( 1 - \frac{x}{16} \right)\,. \label{eq:sol1r1}
\end{align}

Eqs.~\eqref{eq:sol1r1} are apparently different from the solutions determined by studying
the maximal cut~\eqref{eq:cut1} and \eqref{eq:cut2}. Indeed, as the four functions are all solutions 
of the same second order homogeneous differential equation, a relation among them must
exist such that only two functions are really independent.
Indeed, it is well known that the elliptic integral of the first kind $\EK(z)$ and its complementary
$\EK(1-z)$ fulfil the following relations
\begin{align}
&\EK\left( z\right) = \frac{1}{\sqrt{z}} \left( \EK\left( \frac{1}{z}\right) -   i\, \EK\left( 1-\frac{1}{z}\right) \right)  \label{eq:relell1} \\
&\EK\left( 1-z\right) = \frac{1}{\sqrt{z}}  \EK\left( 1-\frac{1}{z}\right)
\label{eq:relell2}
\end{align}
where we use the prescription 
$z \to z + i\, 0^+$ for definiteness.\footnote{
Note that Eqs.~\eqref{eq:relell1} and \eqref{eq:relell2} are not
the only relations between elliptic integrals of different arguments. 
Other relations are known, which we did 
not find necessary in this context.}
Applying Eq.~\eqref{eq:relell1} twice to the solutions~\eqref{eq:cut1} and \eqref{eq:cut2},
we construct two more equivalent sets of solutions
\begin{align}
\left\{ \EK\left(\frac{16}{x}\right)\,, \;\;  \EK\left( 1- \frac{16}{x}\right) \right\} \qquad 
\mbox{and} \qquad 
\left\{ \sqrt{x}\,  \EK\left(\frac{x}{16}\right)\, ,\; \; \sqrt{x}\,\EK\left( 1- \frac{x}{16}\right) \right\}.
\label{eq:invariantset}
\end{align}
It is simple to check by direct insertion that these functions do indeed solve
our second order differential equation.
This proves the equivalence of the solutions found through the maximal cut with the ones found
by solving directly the differential equation.
The existence of more than one representation for the same solutions in terms of elliptic integrals
of different arguments will turn out to be very important to write down the analytic continuation
of the solution from $0<x<16$ to the whole phase space in terms of explicitly real solutions,
as explained in Appendix~\ref{sec:ancont}.

We can now proceed with the solution of the system~\eqref{eq:topsysex}
in the region $0<x<16$; here the most convenient representation is provided
by the choice~\eqref{eq:sol1r1}, which we will adopt from now on.
In order to determine the second row of the matrix $G(x)$ we
plug~\eqref{eq:sol1r1} into~\eqref{eq:deqmat} and find for consistency that
\begin{align}
I_2^{(0,16)}(x) = -\sqrt{x}\,\EE \left(  \frac{x}{16}  \right)\,, \qquad J_2^{(0,16)}(x) = 
\sqrt{x} \left[ \EE \left( 1-\frac{x}{16} \right) - \EK \left( 1- \frac{x}{16} \right) \right]\,, \label{eq:sol2r1}
\end{align} 
where we introduced the complete elliptic integral of the second kind
\begin{equation}
\EE(z) = \int_0^1 dx\, \frac{\sqrt{1-\,z\,x^2}}{\sqrt{1-x^2}} \,, \quad 
\mbox{for} \quad z \in \mathbb{C} \quad \mbox{and}\quad \Re{(z)} < 1\,. \label{eq:ellint2}
\end{equation}
We write therefore the matrix of the solutions valid for $0<x<16$ as
\begin{align}
G^{(0,16)}(x) = \left(\begin{matrix} I_1^{(0,16)}(x) & J_1^{(0,16)}(x)  \\ I_2^{(0,16)}(x) & J_2^{(0,16)}(x) \end{matrix}\right)\,. \label{eq:solmatr1}
\end{align}
We can now compute the Wronskian of the solutions. We find
\begin{align}
\label{eq:wronskianval}
W^{(0,16)}(x) = \det{ \left( G^{(0,16)}(x) \right) } =
 I_1^{(0,16)}(x) J_2^{(0,16)}(x) - J_1^{(0,16)}(x) I_2^{(0,16)}(x) = \frac{\pi}{2}\,x
 \end{align}
which is a direct consequence of the Legendre identity among the first two complete elliptic integrals
\begin{equation}
\EK(z) \EE(1-z) + \EE(z) \EK(1-z) - \EK(z) \EK(1-z) = \frac{\pi}{2}\,.
\end{equation}

Alternatively, we can determine $W(x)$ also without resorting to the Legendre identity.
The fact that the Wronskian must be a linear function of $x$ can be easily seen taking its derivative as
\begin{equation}
\frac{d}{dx} W(x) = \frac{d}{dx} \det{\left( G(x) \right)} = 
Tr\left( G^{(-1)}(x) B(x) G(x)\right) \det \left( G(x) \right) = \frac{1}{x}  W(x)\,,
\end{equation}
which gives as a solution
\begin{equation}
W(x) = c\, x\,, \label{eq:linwr}
\end{equation}
with $c$ an arbitrary integration constant.\footnote{Note that this is of course a general property
of the solution and remains true
independent of the region $a<x<b$ that we consider.}
The value of the constant $c$ can be then fixed
by computing the Wronskian for a fixed value of $x$.
For example, we can study the behaviour of the solutions at the two boundaries, i.e.\
$x \to 0^+$ and $x \to 16^-$. We find respectively, keeping only the leading behaviour,

\begin{align}
\lim_{x \to 0^+} \; I_1^{(0,16)}(x) &= \frac{\pi}{2} \sqrt{x} \,, \nonumber\\
\lim_{x \to 0^+} \; I_2^{(0,16)}(x) &= - \frac{\pi}{2} \sqrt{x} \,,  \nonumber \\
\lim_{x \to 0^+} \; J_1^{(0,16)}(x) &= \frac{\sqrt{x}}{2}\left( -\ln{(x)} + 8\, \ln{(2)} \right)\,, \nonumber \\
\lim_{x \to 0^+} \; J_2^{(0,16)}(x) &= \frac{\sqrt{x}}{2}\left( \ln{(x)} - 8\, \ln{(2)} + 2 \right) \,,\label{eq:expr1a}
\end{align}
and
\begin{align}
\lim_{x \to 16^-} \; I_1^{(0,16)}(x) &= -2\ln{(16-x)} + 16 \ln{(2)} 
+ \frac{(16-x)}{32} \left( \ln{(16-x)} - 8 \ln{(2)} - 2\right) \,,\nonumber \\
\lim_{x \to 16^-} \; I_2^{(0,16)}(x) &=  -4 
+ \frac{(16-x)}{16} \left( \ln{(16-x)} - 8 \ln{(2)} + 3\right)\,,\nonumber \\
\lim_{x \to 16^-} \; J_1^{(0,16)}(x) &= 2\,\pi  - \frac{\pi}{32} (16-x) \,,\nonumber \\
\lim_{x \to 16^-} \; J_2^{(0,16)}(x) &= - \frac{\pi}{16} (16-x) \,. \label{eq:expr1b}
\end{align}
By using~\eqref{eq:expr1a} and~\eqref{eq:expr1b} it is easy to verify explicitly that
\begin{align}
\lim_{x \to 0^+} \frac{W^{(0,16)}(x)}{x} = \lim_{x \to 16^-} \frac{W^{(0,16)}(x)}{x} = \frac{\pi}{2}\,,
\end{align}
which fixes the constant $c$.

\subsection{Analytic continuation of the inhomogeneous term}

With the results of the previous sections we can solve the inhomogeneous
differential equations for the top topology using Euler's
variation of constants.
In order to employ explicitly real building blocks, we need to analytically
continue the homogeneous solutions and the inhomogeneous term to the various
regions of the phase space.
Details for the continuation of the homogeneous solutions are worked out
in Appendix~\ref{sec:ancont}.
For the analytic continuation of the inhomogeneous term we give the explicit
results in this section to highlight the different emerging functions.

For clarity, in every region $a<x<b$ we
extract explicitly the imaginary part (whenever there is one) and
introduce the functions $\R^{(a,b)}(x)$ and  $\I^{(a,b)}(x)$ 
respectively for the real and imaginary parts of $N_{11}^{(4)}(x)$ in that region
$$N_{11}^{(4)}(x) \Big|_{a<x<b} = \R^{(a,b)}(x) + i\, \I^{(a,b)}(x)\,.$$
For the analytical continuation to other kinematical regions we assign an
infinitesimal positive imaginary part to $s$ which translates to a negative
imaginary part for $x$, $x \to x - i\,0^+$.
We will also make use of letters from the alphabet
\begin{align}
l'_i &=  \{ \sqrt{-x}, \tfrac{1}{2}(\sqrt{-x} + \sqrt{-x-4}), \sqrt{-x - 4},
\tfrac{1}{2}(\sqrt{-x} + \sqrt{-x + 4}), \sqrt{-x + 4} \},
\end{align}
depending on the region of phase space.

For the region $4<x<16$ we obtain
\begin{align}
 \R^{(4,16)}(x) &= 5 \ln^2 (l_2)
- 3\, l_1\; \frac{ \zeta_2/2 
+  \ln^2( l_4) +  \Li_2( -1/l_4^2 )}{ l_5}\,,
\nonumber\\
\I^{(4,16)}(x) &= 0\,. \label{eq:R416}
\end{align}
We remark that $x=16$ is actually no special point for the subtopologies
and a single analytic expression is sufficient for the entire range $x>4$.
%\newline
In the region $0<x<4$, $N_{11}^{(4)}(x)$ stays real and we obtain
\begin{align}
\R^{(0,4)}(x) &= 5 \ln^2 (l_2)
+ 3\,l_1\; \frac{\Cl_2(-2 \arccsc(2/l_1) )}{ l_5'}\,,
\nonumber\\
\I^{(0,4)}(x) &=0\,, \label{eq:R04}
\end{align}
where the Clausen function
$
\Cl_2(\theta) = - \int_0^\theta \ln{\left| 2 \sin{\frac{t}{2}}\right|}\, dt
$
arises from the dilogarithm of a pure phase factor,
$\Cl_2(\theta) = \Im\left(\Li_2(e^{i \theta})\right)$,
which can be seen from $l_4^2 = (\sqrt{x}-i\sqrt{4-x})/(\sqrt{x}+i\sqrt{4-x})$.
%\newline
In the region $-4 < x < 0$, the inhomogeneous terms develops an imaginary part and can be
conveniently expressed in terms of
\begin{align}
\R^{(-4,0)}(x) &= - 5\arcsec^2{(2/l_3)} + 3\, l_1'\, \frac{\left(   \zeta_2  -\ln^2{(l_4')}   - 
\Li_2(1/l_4'^2) \right)}{l_5'}\,,
\nonumber\\
\I^{(-4,0)}(x) &= \pi\, \frac{3\, l_1'}{l_5'} \ln{(l_4')}\,, \label{eq:Rm40}
\end{align}
where the first term of the real part arises from a purely imaginary $\ln(l_2)$
since $l_2^2 = (\sqrt{-x} + i \sqrt{4 + x})/(\sqrt{-x} - i \sqrt{4 + x})$
is a pure phase factor.
%\newline
Finally, for $-\infty < x < -4$ we obtain for the real and the imaginary part,
\begin{align}
\R^{(-\infty,-4)}(x) &= 5\ln^2{(l_2')} - \frac{15}{2}\zeta_2 + 3\, l_1'\, \frac{\left(  \zeta_2 - \ln^2{(l_4')}   - \, \Li_2(1/l_4'^2) \right)}{l_5'}\,,
\nonumber\\
\I^{(-\infty,-4)}(x) &= \pi \,\left( \frac{3\, l_1'}{l_5'} \ln{(l_4')}- 5 \, \ln{(l_2')} \right)\,, \label{eq:Rmoom4}
\end{align}
respectively.

\subsection{The inhomogeneous solution}\label{sec:inhomsol}

We are now finally in the position of writing down the complete solution
of~\eqref{eq:topsys} in all relevant regions of the phase space.
For definiteness, we start in the Euclidean region $0<x<4$, $-4 m^2 < s < 0$, 
and then use the results of the previous section and of Appendix~\ref{sec:ancont} to continue the solution
to the other regions.
In Appendix~\ref{sec:ancont}, in particular, we build up matrices of solutions
 $$G^{(a,b)}(x) = 
 \left( \begin{matrix} I_1^{(a,b)}(x) && J_1^{(a,b)}(x) \\ 
 I_2^{(a,b)}(x) && J_2^{(a,b)}(x) \end{matrix}\right)$$
whose entries are explicitly real for $a<x<b$, and we show explicitly how to match 
the homogeneous solution moving from one region to the other.
We recall that the expansions of $m_{10}$ and $m_{11}$ start at order $\e^4$.
We first write down the solutions for $\tm_{10}^{(4)}$ and $\tm_{11}^{(4)}$ and then rotate
them back to obtain expressions for the physical master integrals $m_{10}$ and $m_{11}$.

\subsubsection*{Region $\mathbf{0<x<16}$}

From~\eqref{eq:toprot}, \eqref{eq:invG}, \eqref{eq:wronskianval}
and the inhomogeneous term in~\eqref{eq:nonhomexp}
we get in the region $0<x<4$
\begin{align}
\frac{d}{dx}\,\left( \begin{matrix} \tm_{10}^{(4)} \\ \tm_{11}^{(4)} \end{matrix}\right) = 
\frac{2}{\pi} \frac{1}{x}\,\left( \begin{matrix} J_2^{(0,16)}(x) && -J_1^{(0,16)}(x) \\ -I_2^{(0,16)}(x) && I_1^{(0,16)}(x) \end{matrix}\right)
\left( \begin{matrix} 0 \\  \R^{(0,4)}(x) \end{matrix}\right)
\end{align}
which gives
\begin{align}
&\tm_{10}^{(4)}(x) = - \frac{2}{\pi}\,\int_{0}^{x} \, \frac{dy}{y}\, J_1^{(0,16)}(y)\, \R^{(0,4)}(y) + c_{10}
\,, \nonumber\\
&\tm_{11}^{(4)}(x) =  + \frac{2}{\pi}\, \int_{0}^x \, \frac{dy}{y}\, I_1^{(0,16)}(y)\,  \R^{(0,4)}(y) + c_{11}\,,
\label{eq:solmtilde}
\end{align}
where $c_{10}$ and $c_{11}$ are two integration constants.
We stress again that $\R^{(0,4)}(y)$ is a simple combination of weight two
polylogarithms.
Since neither the top topology nor the subtopologies develop
an imaginary part when $x$ crosses the pseudo-threshold $x=4$, 
we can write an explicitly real solution for the physical (unrotated) master integrals 
$m_{10}$, $m_{11}$
valid for the larger range $0<x<16$ 
\begin{align}
m_{10}^{(4)}(x) &=  c_{10}\, I_1^{(0,16)}(x) + c_{11}\, J_1^{(0,16)}(x) \nonumber \\
&+\frac{2}{\pi} \, J_1^{(0,16)}(x)\, \int_{0}^x \,\frac{dy}{y}\, I_1^{(0,16)}(y)\,\R^{(a,b)}(y)
- \frac{2}{\pi} \, I_1^{(0,16)}(x)\, \int_{0}^x \, \frac{dy}{y}\, J_1^{(0,16)}(y)\, \R^{(a,b)}(y)
\,, \nonumber\\ & \nonumber \\
m_{11}^{(4)}(x) &=  c_{10}\, I_2^{(0,16)}(x) + c_{11}\, J_2^{(0,16)}(x)\,, \nonumber \\
&+\frac{2}{\pi} \, J_2^{(0,16)}(x)\, \int_{0}^x \, \frac{dy}{y}\, I_1^{(0,16)}(y)\, \R^{(a,b)}(y)
- \frac{2}{\pi} \, I_2^{(0,16)}(x)\, \int_{0}^x \, \frac{dy}{y}\, J_1^{(0,16)}(y)\, \R^{(a,b)}(y)\,,
\label{eq:soltofix}
\end{align}
where $(a,b)$ is either $(0,4)$ or $(4,16)$, depending on whether $y<4$ or $y>4$. 
The integration constants must be fixed with two proper boundary conditions.
Imposing that the integrals be regular as $x \to 16^-$ 
and go to zero for $x \to 0^+$, we find $c_{10}=c_{11}=0$
\begin{align}
m_{10}^{(4)}(x) &=  \frac{2}{\pi} \, \left[ J_1^{(0,16)}(x)\, \int_{0}^x \, \frac{dy}{y}\, I_1^{(0,16)}(y)\, 
\R^{(a,b)}(y)
- I_1^{(0,16)}(x)\, \int_{0}^x \, \frac{dy}{y}\, J_1^{(0,16)}(y)\, \R^{(a,b)}(y)
\right]
\,, \nonumber\\ & \nonumber \\
m_{11}^{(4)}(x) &=  
\frac{2}{\pi} \, \left[ J_2^{(0,16)}(x)\, \int_{0}^x \, \frac{dy}{y}\, I_1^{(0,16)}(y)\, \R^{(a,b)}(y)
- I_2^{(0,16)}(x)\, \int_{0}^x \, \frac{dy}{y}\, J_1^{(0,16)}(y)\, \R^{(a,b)}(y) \right] \,.
\label{eq:solfixed}
\end{align}

\subsubsection*{Region $\mathbf{-4<x<0}$}

We continue the solution to the region $-4<x<0$ using $x \to x - i\,0^+$,
\eqref{eq:Rm40}, \eqref{eq:matching} and~\eqref{eq:mxmatch} and obtain
\begin{align}
m_{10}^{(4)}(x) =   \frac{2}{\pi} \, 
&\left[ J_1^{(-\infty,0)}(x)\, \int_{0}^x \, \frac{dy}{y}\, I_1^{(-\infty,0)}(y)\,  \R^{(-4,0)}(y)
  -  \, I_1^{(-\infty,0)}(x)\, \int_{0}^x \, \frac{dy}{y}\, J_1^{(-\infty,0)}(y)\,  \R^{(-4,0)}(y) \right]\nonumber \\
+ i\, \frac{2}{\pi}\, 
&\left[  J_1^{(-\infty,0)}(x)\, \int_{0}^x \, \frac{dy}{y}\, I_1^{(-\infty,0)}(y)\, \I^{(-4,0)}(y)
-  I_1^{(-\infty,0)}(x)\, \int_{0}^x \, \frac{dy}{y}\, J_1^{(-\infty,0)}(y) \, \, \I^{(-4,0)}(y)\right]
\,, \nonumber\\  & \nonumber \\
%%%%
%%%%
m_{11}^{(4)}(x) =   \frac{2}{\pi} \, 
&\left[ J_2^{(-\infty,0)}(x)\, \int_{0}^x \, \frac{dy}{y}\, I_1^{(-\infty,0)}(y)\, \R^{(-4,0)}(y)
-  \, I_2^{(-\infty,0)}(x)\, \int_{0}^x \, \frac{dy}{y}\, J_1^{(-\infty,0)}(y)\, \R^{(-4,0)}(y) \right]\nonumber \\
+ i\, \frac{2}{\pi}\, 
&\left[  J_2^{(-\infty,0)}(x)\, \int_{0}^x \, \frac{dy}{y}\, I_1^{(-\infty,0)}(y)\, \I^{(-4,0)}(y)
-  I_2^{(-\infty,0)}(x)\, \int_{0}^x \, \frac{dy}{y}\, J_1^{(-\infty,0)}(y) \, \, \I^{(-4,0)}(y)\right]\,.
\label{eq:solmink1}
\end{align}
Note that now all integrals are explicitly real and all imaginary parts are explicit.

\subsubsection*{Region $\mathbf{-\infty<x<-4}$}
We can then go further and continue beyond the threshold for the production of two massive
particles, i.e.\ $s>4 m^2$, $x<-4$. Using again formulas~\eqref{eq:Rmoom4},~\eqref{eq:matching} and~\eqref{eq:mxmatch}
we obtain
\begin{align}
m_{10}^{(4)}(x) =    \frac{2}{\pi} \, 
&\left[ J_1^{(-\infty,0)}(x)\, \left( \int_{-4}^x \, \frac{dy}{y}\, I_1^{(-\infty,0)}(y)\,  \R^{(-\infty,-4)}(y) - 
 c_1 \right) \right. \nonumber \\
  -  \, &\;\left. I_1^{(-\infty,0)}(x)\, \left( \int_{-4}^x \, \frac{dy}{y}\, J_1^{(-\infty,0)}(y)\,  \R^{(-\infty,-4)}(y)
  -  c_2 \right) \right]\nonumber \\
+ i\, \frac{2}{\pi}\, 
&\left[  J_1^{(-\infty,0)}(x)\,\left( \int_{-4}^x \, \frac{dy}{y}\, I_1^{(-\infty,0)}(y)\,  \I^{(-\infty,-4)}(y) - 
 c_3 \right) \right. \nonumber \\
  -  \, &\;\left. I_1^{(-\infty,0)}(x)\, \left( \int_{-4}^x \, \frac{dy}{y}\, J_1^{(-\infty,0)}(y)\,  \I^{(-\infty,-4)}(y)
  -  c_4\right)\right]
\,, 
\nonumber\\
%%%%
m_{11}^{(4)}(x) =    \frac{2}{\pi} \, 
&\left[ J_2^{(-\infty,0)}(x)\, \left( \int_{-4}^x \, \frac{dy}{y}\, I_1^{(-\infty,0)}(y)\,  \R^{(-\infty,-4)}(y) - 
 c_1 \right) \right. \nonumber \\
  -  \, &\;\left. I_2^{(-\infty,0)}(x)\, \left( \int_{-4}^x \, \frac{dy}{y}\, J_1^{(-\infty,0)}(y)\,  \R^{(-\infty,-4)}(y)
  -  c_2 \right) \right]\nonumber \\
+ i\, \frac{2}{\pi}\, 
&\left[  J_2^{(-\infty,0)}(x)\,\left( \int_{-4}^x \, \frac{dy}{y}\, I_1^{(-\infty,0)}(y)\,  \I^{(-\infty,-4)}(y) - 
 c_3 \right) \right. \nonumber \\
  -  \, &\;\left. I_2^{(-\infty,0)}(x)\, \left( \int_{-4}^x \, \frac{dy}{y}\, J_1^{(-\infty,0)}(y)\,  \I^{(-\infty,-4)}(y)
  -  c_4\right)\right]
\,, 
\label{eq:solmink2}
\end{align}
where all imaginary parts are explicit and all integrals are real.
The matching constants are
\begin{align}
&c_1 = \int_{-4}^0 \, \frac{dy}{y}\, I_1^{(-\infty,0)}(y)\,  \R^{(-4,0)}(y)\,, \quad 
c_2 = \int_{-4}^0 \, \frac{dy}{y}\, J_1^{(-\infty,0)}(y)\,  \R^{(-4,0)}(y) \,, \nonumber \\
&c_3 = \int_{-4}^0 \, \frac{dy}{y}\, I_1^{(-\infty,0)}(y)\,  \I^{(-4,0)}(y) \,, \quad
c_4 =  \int_{-4}^0 \, \frac{dy}{y}\, J_1^{(-\infty,0)}(y)\,  \I^{(-4,0)}(y) \,.
\end{align}
They can be evaluated with high precision starting
directly from this integral representation.

\subsubsection*{Numerical evaluation}

\begin{figure}[ht]
\centering
  \includegraphics[width=\linewidth]{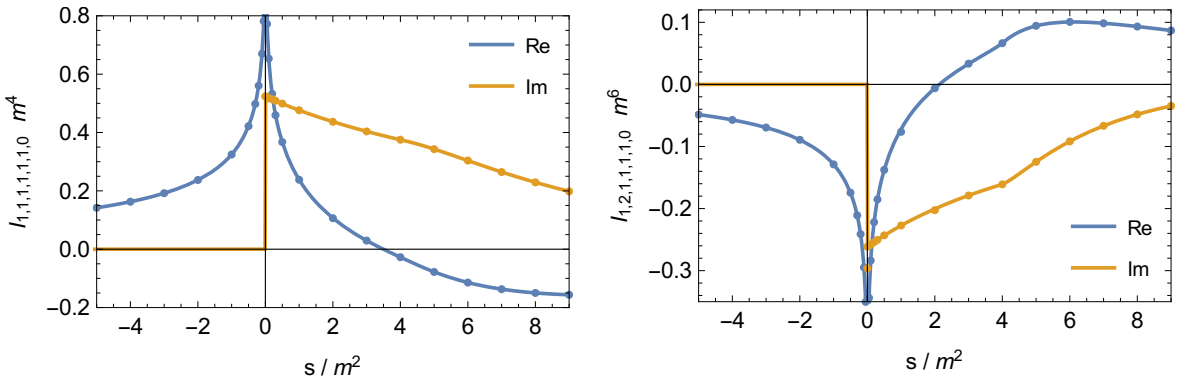}
  \caption{\label{fig:i1x11110}
  Finite top-level master integrals $m^4\, I_{1,1,1,1,1,1,0} $ (left)
  and $m^6\, I_{1,2,1,1,1,1,0} $ (right).
  Solid curves were obtained with our representations,
  dots with SecDec\;3.
  }
\end{figure}

We verified our formulas in the previous section with a thorough comparison
against SecDec\;3, both in the
physical and in the non-physical region.
Numerical results for the top-level master integrals are shown in Figure~\ref{fig:i1x11110}.
We would like to point out that our representation allows for particularly fast
and precise evaluations also in the physical region of phase space.
As an example for a result with higher precision, we obtain
\begin{align}
\left. m^4 I_{1,1,1,1,1,1,0}\right|_{s=5m^2, d=4} &\approx 
-0.07776462028160023644086669458011467822536257409024
\nonumber\\
&~+ i~ 0.34306740464518688969054397597465622650767181505054
\end{align} 
for one of our six-line master integrals.

\section{Conclusions} \label{sec:conclusions} 

Feynman integrals which evaluate to classes of functions outside the realm of 
multiple polylogarithms
constitute the bottleneck for many multiloop calculations relevant for LHC phenomenology.
To this day, only a very limited number of examples have been considered in the literature.
An important step towards a more complete understanding of the new mathematical
structures therefore consists of the study of more explicit examples in order to expose their common properties.
In this paper we have considered the calculation of a two-loop non-planar three-point function
with four massive internal propagators and one external off-shell leg
using the method of differential equations.
The differential equations for the subtopologies can be put in canonical form
and integrated in terms of multiple polylogarithms over an irrational alphabet.
In general, in the presence of different roots, traditional approaches to
integrate the result in terms of Goncharov's polylogarithms fail,
unless one is able to find a change of variables which linearizes
or at least rationalizes the letters of the alphabet.
We employed a new algorithm to perform the integration without any
rationalization of the alphabet.
This allowed us to write the results up to weight four in terms
of simple logarithms, classical polylogarithms and $\Li_{2,2}$ functions only.

We moved then to the two master integrals of the top level sector, which fulfil
two coupled differential equations in the momentum transfer squared. By studying the maximal cut
of the two master integrals we showed how to solve the differential equations
in terms of complete elliptic integrals of first and second kind; we then used Euler's variation
of constants in order to write the solution as a one-fold integral over elliptic integrals and 
multiple polylogarithms.
Finally, we used the same techniques described in~\cite{Remiddi:2016gno} in order to
continue our results to the Minkowski physical region. In this way we provided
expressions which can be evaluated in a fast and reliable manner in the
entire physical region of the phase space.
We compared our results with {SecDec\;3} and found perfect agreement.
We expect that similar techniques can be extended to study also more complicated
three- and four-point functions, as in part confirmed by the results 
recently obtained in~\cite{Bonciani:2016qxi}.

\section*{Acknowledgments}

We would like to thank Stefan Groote, Ettore Remiddi, Robert Schabinger
and Erich Weihs for valuable discussions.
We are grateful to the Mainz Institute for Theoretical Physics (MITP) for its
hospitality and support.

\appendix

\section{Solutions for the subtopologies}\label{app:subtopos}
In this Appendix we report the explicit results for the subtopologies in
the Euclidean region with \mbox{$s<-4m^2$}.
Having a canonical form, we can write the $\epsilon$ expansion through
to weight four in a relatively compact form in terms
of simple functions, including only logarithms, classical polylogarithms and $\Li_{2,2}$ functions
with arguments 
built from the alphabet defined in~\eqref{eq:alphabet}.
Expanding the master integrals according to
\begin{equation}
m_j = \sum_{k=0}^4\, m_j^{(k)}\, \epsilon^k\,+ \mathcal{O}(\epsilon^5)\,.
\end{equation}
we find for the coefficients $m_j^{(k)}$

\input{solutions_subtopos}

\section{Details on the analytic continuation of the homogeneous solution}\label{sec:ancont}
Our goal here is to write down a set of solutions of the homogeneous equation~\eqref{eq:secdeqm}
in terms of real-valued functions for all regions of the phase space. This will allow us to analytically continue the
inhomogeneous solution found in Section~\ref{sec:inhomsol} to the physical region
and extract all imaginary parts explicitly in terms of real functions.
We will consider in order the two relevant regions, i.e.\ the Minkowski region $-\infty < x< 0$ ~($0<s<\infty$)\, 
and the part of the Euclidean region with $16 < x < \infty$ ~($s<-16 m^2$)\,.
Our strategy is to first derive real-valued homogeneous solutions
for each region separately and later relate the different solution
sets using matching matrices.

\subsection*{Minkowski region $\mathbf{-\infty <x<0}$}
In order to obtain a complete set of solutions defined in the Minkowski region we recall that for
any value of $x$ a general solution of the second order differential equation~\eqref{eq:secdeq} 
can always be written as a linear combination of two independent solutions.
Our starting point is \eqref{eq:fsol},
\begin{equation}
f(x) = c_{1}\, \EK \left(  \frac{x}{16} \right) + c_{2} \EK \left( 1 - \frac{x}{16} \right)\,.
\end{equation}
In the Minkowski region we have $s > 0$ with $s \to s + i 0^+$, i.e.\ $x < 0$ with $x \to x - i \, 0^+$ 
and we can take as two independent real solutions
\begin{align}
I_1^{(-\infty,0)}(x) = -\sqrt{-x}\, \EK \left(  \frac{x}{16} \right)\,, \qquad 
J_1^{(-\infty,0)}(x) = \sqrt{-x}\, \left[ \EK \left( 1 - \frac{x}{16} \right) - i\, \EK\left( \frac{x}{16} \right) \right] \, \label{eq:sol1r2v1}
\end{align}
where the sign of the imaginary part is fixed requiring that for $x \to x - i \, 0^+$\, the solution
is real for $-\infty < x < 0$. 
As for the Euclidean region discussed in Section~\ref{sec:toptopo}, we can determine the two missing solutions 
by consistency
\begin{align}
I_2^{(-\infty,0)}(x) =  \sqrt{-x}\, \EE\left( \frac{x}{16} \right)\,, \quad  \;
J_2^{(-\infty,0)}(x) = \sqrt{-x}\, \left[ \EE\left( 1- \frac{x}{16} \right) - 
\EK\left( 1- \frac{x}{16} \right) + i\, \EE\left( \frac{x}{16}  \right) \right]\,.
\end{align}
This solution is not optimal, since it is not explicitly real for $-\infty<x<0$. 
To find a real valued representation of the solution we can use the other set of functions
found in~\eqref{eq:invariantset}.
We find in particular that for $-\infty<x<0$ the follow identities hold identically
\begin{align}
\sqrt{-x}\, \left[ \EK \left( 1 - \frac{x}{16} \right) - i\, \EK\left( \frac{x}{16} \right) \right] &= 
4\, \EK \left( \frac{16}{x}\right)\nonumber \\
\sqrt{-x}\, \left[ \EE\left( 1- \frac{x}{16} \right) - 
\EK\left( 1- \frac{x}{16} \right) + i\, \EE\left( \frac{x}{16}  \right) \right] &=
\frac{x-16}{4} \EK\left( \frac{16}{x}\right) - \frac{x}{4} \EE \left( \frac{16}{x} \right)\,,
\end{align}
such that a basis of functions explicitly real in the whole domain $-\infty<x<0$ can be chosen as
\begin{align}
I_1^{(-\infty,0)}(x) &= -\sqrt{-x}\, \EK \left(  \frac{x}{16} \right)\,, &
J_1^{(-\infty,0)}(x) &= 4\, \EK \left( \frac{16}{x}\right) \,,  \nonumber\\
I_2^{(-\infty,0)}(x) &= \sqrt{-x}\, \EE\left( \frac{x}{16} \right)\,, &
J_2^{(-\infty,0)}(x) &=  \frac{x-16}{4} \EK\left( \frac{16}{x}\right) - \frac{x}{4} \EE \left( \frac{16}{x} \right) \,.
\label{eq:sol1r2}
\end{align}

Similar to our analysis for the region $0<x<16$, 
it is useful to study the limiting behaviour of these functions close to the two boundaries
\begin{align}
\lim_{x \to -\infty} \; I_1^{(-\infty,0)}(x) &=  -2\ln{(-x)} - \frac{8 \ln{(-x)} - 16}{x}\,, \nonumber \\
\lim_{x \to -\infty} \; I_2^{(-\infty,0)}(x) &=  -\frac{x}{4} - \ln{(-x)} + 1 - \frac{2 \ln{(-x)} - 3}{x}\nonumber \\
\lim_{x \to -\infty} \; J_1^{(-\infty,0)}(x) &= 2\pi + \frac{8 \pi}{x}\,, \nonumber\\
\lim_{x \to -\infty} \; J_2^{(-\infty,0)}(x) &= -\pi - \frac{2 \pi}{x} \label{eq:expr2a}
\end{align}
and
\begin{align}
\lim_{x \to 0^-} \; I_1^{(-\infty,0)}(x) &= -\frac{\pi}{2} \sqrt{-x} \,,\nonumber\\
\lim_{x \to 0^-} \; I_2^{(-\infty,0)}(x) &= \frac{\pi}{2} \sqrt{-x} \nonumber \\
\lim_{x \to 0^-} \; J_1^{(-\infty,0)}(x) &= \frac{\sqrt{-x}}{2}\left( -\ln{(-x)} + 8\, \ln{(2)} \right) \,, \nonumber \\
 \lim_{x \to 0^-} \; J_2^{(-\infty,0)}(x) &= \frac{\sqrt{-x}}{2}\left( \ln{(-x)} - 8\, \ln{(2)} + 2 \right) \,. \label{eq:expr2b}
\end{align}

As discussed above, the Wronskian of the solutions must be a linear function of $x$, see~\eqref{eq:linwr}.
It is simple starting from~\eqref{eq:expr2a} and~\eqref{eq:expr2b}
to prove that we have
\begin{align}
\lim_{x \to -\infty} \frac{W^{(-\infty,0)}(x)}{x} = \lim_{x \to 0^-} \frac{W^{(-\infty,0)}(x)}{x} = \frac{\pi}{2}
\end{align}
also for our solutions in this region.

\subsection*{Euclidean region $\mathbf{16 < x < \infty}$}
The same idea can be applied in the remaining part of the Euclidean region. Similar to the Minkowski region, we start here with solutions constructed as
complex linear combinations of the original functions which evaluate to real numbers for $16<x<\infty$,
\begin{align}
I_1^{(16,\infty)}(x) &= \sqrt{x}\,\left [ \EK \left(  \frac{x}{16} \right)  + i\, \EK\left( 1- \frac{x}{16} \right) \right]\,, \nonumber\\
J_1^{(16,\infty)}(x) &= \sqrt{x}\, \EK \left( 1 - \frac{x}{16} \right) \, \label{eq:sol1r3v1}
\intertext{and}
I_2^{(16,\infty)}(x) &= - \sqrt{x}\, \left[ \EE\left( \frac{x}{16} \right) - i\, \left(  \EE \left( 1- \frac{x}{16} \right) 
- \EK\left( 1- \frac{x}{16} \right) \right) \right]\,, \nonumber \\
J_2^{(16,\infty)}(x) &= \sqrt{x}\, \left[ \EE\left( 1- \frac{x}{16} \right) - \EK\left( 1- \frac{x}{16} \right) \right]\,.
\end{align}
Again we use $x \to x - i 0^+$. Similarly as before, this
representation of the solutions is not optimal since it is not explicitly real. 
Using the sets of solutions~\eqref{eq:invariantset}
we find the following identities, valid for $16 < x < \infty$ with $x \to x -i 0^+$,
\begin{align}
\sqrt{x}\,\left [ \EK \left(  \frac{x}{16} \right)  + i\, \EK\left( 1- \frac{x}{16} \right) \right] &= 4\, \EK \left( \frac{16}{x}\right)\,,
\nonumber \\
- \sqrt{x}\, \left[ \EE\left( \frac{x}{16} \right) - i\, \left(  \EE\left( 1- \frac{x}{16} \right) 
- \EK\left( 1- \frac{x}{16} \right) \right) \right] &= 
\frac{x-16}{4} \EK\left( \frac{16}{x}\right) - \frac{x}{4} \EE \left( \frac{16}{x} \right) \,.
\end{align}
A possible choice of real solutions for the region $16 < x < \infty$ is therefore
\begin{align}
I_1^{(16,\infty)}(x) &=  4\, \EK \left( \frac{16}{x}\right) \,, &
J_1^{(16,\infty)}(x) &= \sqrt{x}\, \EK \left( 1 - \frac{x}{16} \right) \, \nonumber \\ 
I_2^{(16,\infty)}(x) &=\frac{x-16}{4} \EK\left( \frac{16}{x}\right) - \frac{x}{4} \EE \left( \frac{16}{x} \right) \,,  &
J_2^{(16,\infty)}(x) &= \sqrt{x}\, \left[ \EE\left( 1- \frac{x}{16} \right) - \EK\left( 1- \frac{x}{16} \right) \right]\,.
\label{eq:sol1r3}
\end{align}

Also in this last case we  shall study the limiting values at the two boundaries $x \to 16^+$ and
$x \to + \infty$. We find
\begin{align}
\lim_{x \to 16^+} \; I_1^{(16,\infty)}(x) &= -2\ln{(16-x)} + 16 \ln{(2)} 
+ \frac{(16-x)}{32} \left( \ln{(16-x)} - 8 \ln{(2)} - 2\right) \,,\nonumber \\
\lim_{x \to 16^+} \; I_2^{(16,\infty)}(x) &=  -4 
+ \frac{(16-x)}{16} \left( \ln{(16-x)} - 8 \ln{(2)} + 3\right) \,,\nonumber \\
\lim_{x \to 16^+} \; J_1^{(16,\infty)}(x) &= 2\,\pi  - \frac{\pi}{32} (16-x) \,,\nonumber \\ 
\lim_{x \to 16^+} \; J_2^{(16,\infty)}(x) &= - \frac{\pi}{16} (16-x) \,, \label{eq:expr3a}
\intertext{and}
\lim_{x \to +\infty} \; I_1^{(16,\infty)}(x) &=  2\pi + \frac{8 \pi}{x}\,,\nonumber \\
\lim_{x \to +\infty} \; I_2^{(16,\infty)}(x) &=  -\pi - \frac{2 \pi}{x}\,, \nonumber \\
\lim_{x \to +\infty} \; J_1^{(16,\infty)}(x) &= 2\ln{(x)} + \frac{8 \ln{(x)} - 16}{x}\,, \nonumber \\
\lim_{x \to +\infty} \; J_2^{(16,\infty)}(x) &= \frac{x}{4} - \ln{(x)} - 1 - \frac{2 \ln{(x)} - 3}{x}\,.\label{eq:expr3b}
\end{align}

Finally, we compute the value of the Wronskian also in this part of the Euclidean region.
As before, it must be a linear function of $x$, see~\eqref{eq:linwr}.
Again, starting from the limiting behaviours above~\eqref{eq:expr3a} and~\eqref{eq:expr3b}, 
we get at once
\begin{align}
\lim_{x \to 16^+} \frac{W^{(16,\infty)}(x)}{x} = \lim_{x \to +\infty} \frac{W^{(16,\infty)}(x)}{x} = \frac{\pi}{2}\,.
\end{align}

We stress once more that of course the overall normalization of the Wronskian depends on the
overall normalization of the functions chosen as a solution in the given region. 
It is useful for simplicity to fix it such that the Wronskian assumes always the same value in
every region.

\subsection*{The matching}
In this section we will show how to match the solutions found in the three different regions, by
analytically continuing them across the three boundaries $x=0$, $x=16$ and $x=\pm \infty$.
Given the matrix of the solutions defined in the region $a<x<b$,
\[
G^{(a,b)}(x) = \left(\begin{matrix} I_1^{(a,b)}(x) & J_1^{(a,b)}(x)  \\ I_2^{(a,b)}(x) & J_2^{(a,b)}(x) \end{matrix}\right)\,,
\]
we define the matching matrix
in the point $x=b$, $M^{(b)}$, which allows to continue the solution to the next region $b<x<c$ as follows
\begin{equation}
G^{(b,c)}(x) = G^{(a,b)}(x) M^{(b)}\,. \label{eq:matching}
\end{equation}
We perform the matching analytically at each of the three points,
where we implement Feynman's causality prescription by adding to $x$
a small negative imaginary part, $x \to x - i\,0^+$.
The matching is straightforward at this point, since we derived all required
limits already in Eqs.~\eqref{eq:expr1a}, \eqref{eq:expr1b}, \eqref{eq:expr2a}, \eqref{eq:expr2b}, 
\eqref{eq:expr3a} and \eqref{eq:expr3b}. The matching matrices read
\begin{align}
M^{\infty} &=  \left(\begin{matrix} -i && 1  \\ -1 && 0\end{matrix}\right) \,, & 
M^{0} &= \left(\begin{matrix} i & -1  \\ 0 & -i\end{matrix}\right)\,, & 
M^{16} &= \left(\begin{matrix} 1 && 0  \\ i && 1\end{matrix}\right)\,. \label{eq:mxmatch}
\end{align}
As expected we find consistently that
$$M^\infty \, M^0\, M^{16} = M^{16} M^{\infty} M^{0} = M^{0} M^{16} M^{\infty} =  
\left(\begin{matrix} 1 && 0  \\ 0 && 1\end{matrix}\right)\,.$$

As an example, in order to continue the solution valid in the range $0<x<16$ 
to the Minkowski region $-\infty < x< 0$, we must continue across $x=0$
through the matrix $M^{0}$ as follows
\begin{align}
G^{(0,16)}(x) = G^{(-\infty,0)}(x)\, M^{0} 
\end{align}
i.e.\
\begin{align}
  & \left(\begin{matrix} I_1^{(0,16)}(x) && J_1^{(0,16)}(x)  \\ I_2^{(0,16)}(x) && J_2^{(0,16)}(x) \end{matrix}\right)
  =  
  \left(\begin{matrix} I_1^{(-\infty,0)}(x) && J_1^{(-\infty,0)}(x)  \\ I_2^{(-\infty,0)}(x) && 
  J_2^{(-\infty,0)}(x) \end{matrix}\right) \left(\begin{matrix} i & -1  \\ 0 & -i\end{matrix}\right) \nonumber \\
\end{align}
which gives, for $x<0$, 
\begin{align}
&I_1^{(0,16)}(x) \;\; \longrightarrow \;\;  i\, I_1^{(-\infty,0)}(x)\,, \nonumber \\
&J_1^{(0,16)}(x) \;\; \longrightarrow \;\; -I_1^{(-\infty,0)}(x) - i\, J_1^{(-\infty,0)}(x)\,, \nonumber \\
&I_2^{(0,16)}(x) \;\; \longrightarrow  \;\; i\, I_2^{(-\infty,0)}(x)\,, \nonumber \\
&J_2^{(0,16)}(x) \;\; \longrightarrow \;\; -I_2^{(-\infty,0)}(x) - i\, J_2^{(-\infty,0)}(x) \,.
\end{align}

\newpage 
\bibliographystyle{JHEP}   
\bibliography{Biblio}

\end{document}

%% file: solutions_subtopos.tex
\begin{align}
m_1^{(0)} &= 1\,, \nonumber \\
m_1^{(j)} &= 0\,, \quad \text{for~} j \ge 1,
\\[4ex]
m_2^{(0)} &= -1 \,, \nonumber \\
m_2^{(1)} &= 2 \log ( l_1 ) \,, \nonumber \\
m_2^{(2)} &= \frac{\pi ^2}{6}-2 \log ^2(l_1)\,, \nonumber \\
m_2^{(3)} &= \frac{4 \log ^3(l_1)}{3}-\frac{1}{3} \pi ^2 \log (l_1)+2 \zeta_3\,,  \nonumber \\
m_2^{(4)} &= -4 \zeta_3 \log (l_1)-\frac{2}{3} \log ^4(l_1)+\frac{1}{3} \pi ^2 
\log^2(l_1)+\frac{\pi ^4}{40}\,,
\\[4ex]
m_3^{(0)} &= 0 \,,\quad
m_3^{(1)} = 0 \,, \nonumber \\
m_3^{(2)} &= 4 \log ^2(l_2) \,, \nonumber \\
m_3^{(3)} &= 8 \log (l_1) \log ^2(l_2)-12 \text{Li}_3\left(\frac{1}{l_2^2}\right)-16
   \text{Li}_2\left(\frac{1}{l_2^2}\right) \log (l_2)+24
   \text{Li}_3\left(\frac{1}{l_2 l_3}\right)\nonumber \\ &+24
   \text{Li}_3\left(\frac{l_2}{l_3}\right)+24 \log (l_2)
   \text{Li}_2\left(\frac{1}{l_2 l_3}\right)+12 \log (l_2) \log
   ^2(l_3)-20 \log ^3(l_2)-\frac{10}{3} \pi ^2 \log (l_2)\nonumber \\&-8 \log
   ^3(l_3)+4 \pi ^2 \log (l_3)-30 \zeta_3\,, \nonumber \\
   %%%%%%
m_3^{(4)} &= -24 \text{Li}_{2,2}\left(-1,-\frac{1}{l_2^2}\right)-48 \log (l_2)
   \text{Li}_3\left(-\frac{1}{l_1 l_2^2 l_3}\right)+32 \log (l_1)
   \text{Li}_2\left(\frac{1}{l_2^2}\right) \log (l_2)\nonumber \\
   &-48 \log (l_1)
   \log (l_2) \text{Li}_2\left(\frac{1}{l_2 l_3}\right)+48 \log
   (l_1) \text{Li}_3\left(\frac{1}{l_2 l_3}\right)+48 \log (l_1)
   \text{Li}_3\left(\frac{l_2}{l_3}\right)\nonumber \\&
   +24 \log ^2(l_1) \log
   (l_2) \log (l_3)-24 \text{Li}_4\left(-\frac{1}{l_1
   l_2}\right)+24 \text{Li}_4\left(\frac{l_1}{l_2}\right)-32 \log
   (l_2) \text{Li}_3\left(\frac{l_1}{l_2}\right)
   \nonumber \\&+4 \log ^3(l_1)
   \log (l_2)-76 \log (l_1) \log ^3(l_2)+10 \log ^2(l_1) \log
   ^2(l_2)+\frac{8}{3} \pi ^2 \log (l_1) \log (l_2)
   \nonumber \\&-16 \log (l_1)
   \log ^3(l_3)+8 \pi ^2 \log (l_1) \log (l_3)-84 \zeta_3 \log
   (l_1)-\log ^4(l_1)-2 \pi ^2 \log ^2(l_1)
   \nonumber \\&+48
   \text{Li}_2\left(\frac{1}{l_2^2}\right) \text{Li}_2\left(\frac{1}{l_2
   l_3}\right)+24 \text{Li}_2\left(\frac{1}{l_2^2}\right) \log
   ^2(l_3)+48 \text{Li}_2\left(\frac{1}{l_2^2}\right) \log (l_2) \log
   (l_3)-8 \text{Li}_2\left(\frac{1}{l_2^2}\right){}^2
   \nonumber \\ &-\frac{4}{3} \pi ^2
   \text{Li}_2\left(\frac{1}{l_2^2}\right)+36
   \text{Li}_4\left(-\frac{1}{l_2^2}\right)-12
   \text{Li}_4\left(\frac{1}{l_2^2}\right)-40
   \text{Li}_2\left(\frac{1}{l_2^2}\right) \log ^2(l_2)-112
   \text{Li}_3\left(\frac{1}{l_2^2}\right) \log (l_2)
   \nonumber \\&-36
   \text{Li}_2\left(\frac{1}{l_2 l_3}\right){}^2+4 \pi ^2
   \text{Li}_2\left(\frac{1}{l_2 l_3}\right)+60 \log ^2(l_2)
   \text{Li}_2\left(\frac{1}{l_2 l_3}\right)-36 \log ^2(l_3)
   \text{Li}_2\left(\frac{1}{l_2 l_3}\right)
   \nonumber \\&-72 \log (l_2) \log
   (l_3) \text{Li}_2\left(\frac{1}{l_2 l_3}\right)+24 \log (l_2)
   \text{Li}_3\left(\frac{1}{l_2 l_3}\right)+72 \log (l_2)
   \text{Li}_3\left(\frac{l_2}{l_3}\right)
   \nonumber \\&-44 \log (l_2) \log
   ^3(l_3)+60 \log ^3(l_2) \log (l_3)+66 \log ^2(l_2) \log
   ^2(l_3)+8 \pi ^2 \log (l_2) \log (l_3)
   \nonumber \\&+28 \zeta_3 \log
   (l_2)-\frac{26 \log ^4(l_2)}{3}-\frac{62}{3} \pi ^2 \log ^2(l_2)-9
   \log ^4(l_3)+2 \pi ^2 \log ^2(l_3)-\frac{61 \pi ^4}{180}\,,
\\[4ex]
m_4^{(0)} &= 0 \,, \nonumber \\
m_4^{(1)} &= 2 \log (l_2) \,,\nonumber \\
m_4^{(2)} &= -4 \log (l_1) \log (l_2)+2 \text{Li}_2\left(\frac{1}{l_2^2}\right)-6
   \text{Li}_2\left(\frac{1}{l_2 l_3}\right)-6 \log (l_2) \log
   (l_3)\nonumber \\&
   +5 \log ^2(l_2)-3 \log ^2(l_3)+\frac{\pi ^2}{6}\,, \nonumber \\
m_4^{(3)} &= -6 \text{Li}_3\left(-\frac{1}{l_1 l_2^2 l_3}\right)+12 \log (l_1)
   \log (l_2) \log (l_3)-8
   \text{Li}_3\left(\frac{l_1}{l_2}\right)+6 \log ^2(l_1) \log
   (l_2)\nonumber \\&
   -20 \log (l_1) \log ^2(l_2)+3 \log ^2(l_1) \log
   (l_3)+3 \log (l_1) \log ^2(l_3)+\log ^3(l_1)+\pi ^2 \log
   (l_1)\nonumber \\&
   -16 \text{Li}_3\left(\frac{1}{l_2^2}\right)+24
   \text{Li}_3\left(\frac{l_2}{l_3}\right)+18 \log (l_2) \log
   ^2(l_3)-\frac{20 \log ^3(l_2)}{3}-\frac{16}{3} \pi ^2 \log (l_2)
   \nonumber \\&-3
   \log ^3(l_3)+3 \pi ^2 \log (l_3)-5 \zeta_3\,, \nonumber \\
  % \displaybreak
m_4^{(4)} &= 12 \text{Li}_{2,2}\left(-\frac{1}{l_1
   l_2},\frac{l_1}{l_2}\right)-18 \text{Li}_4\left(-\frac{1}{l_1
   l_2^2 l_3}\right)-18 \text{Li}_4\left(\frac{l_1
   l_3}{l_2^2}\right)-6 \log ^2(l_1)
   \text{Li}_2\left(\frac{1}{l_2^2}\right) 
   \nonumber \\
   &+12 \log (l_1)
   \text{Li}_2\left(\frac{1}{l_2^2}\right) \log (l_2)+48 \log (l_3)
   \text{Li}_3\left(\frac{l_1}{l_2}\right)-18 \log ^2(l_1) \log
   (l_2) \log (l_3)
   \nonumber \\&-18 \log (l_1) \log (l_2) \log
   ^2(l_3)+12 \log (l_1) \log ^2(l_2) \log (l_3)+32
   \text{Li}_4\left(-\frac{1}{l_1 l_2}\right)-4
   \text{Li}_4\left(\frac{l_1}{l_2}\right)
   \nonumber \\&
   -48 \log (l_2)
   \text{Li}_3\left(\frac{l_1}{l_2}\right)+\frac{34}{3} \log ^3(l_1)
   \log (l_2)+44 \log (l_1) \log ^3(l_2)-22 \log ^2(l_1) \log
   ^2(l_2)
   \nonumber \\&+\frac{4}{3} \pi ^2 \log (l_1) \log (l_2)-3 \log
   ^3(l_1) \log (l_3)-3 \log (l_1) \log ^3(l_3)-\frac{9}{2} \log
   ^2(l_1) \log ^2(l_3)
   \nonumber \\&-3 \pi ^2 \log (l_1) \log (l_3)+\frac{7
   \log ^4(l_1)}{12}+\frac{13}{6} \pi ^2 \log ^2(l_1)+24
   \text{Li}_3\left(\frac{1}{l_2^2}\right) \log (l_3)-6
   \text{Li}_2\left(\frac{1}{l_2^2}\right){}^2
   \nonumber \\&-264
   \text{Li}_4\left(-\frac{1}{l_2^2}\right)-164
   \text{Li}_4\left(\frac{1}{l_2^2}\right)-54
   \text{Li}_2\left(\frac{1}{l_2^2}\right) \log ^2(l_2)-120
   \text{Li}_3\left(\frac{1}{l_2^2}\right) \log (l_2)-72
   \text{Li}_4\left(\frac{1}{l_2 l_3}\right)
   \nonumber \\&+72
   \text{Li}_4\left(\frac{l_2}{l_3}\right)+72 \log ^2(l_2)
   \text{Li}_2\left(\frac{1}{l_2 l_3}\right)+144 \log (l_2)
   \text{Li}_3\left(\frac{1}{l_2 l_3}\right)+144 \log (l_2)
   \text{Li}_3\left(\frac{l_2}{l_3}\right)
   \nonumber \\&-78 \log (l_2) \log
   ^3(l_3)+184 \log ^3(l_2) \log (l_3)+18 \log ^2(l_2) \log
   ^2(l_3)+38 \pi ^2 \log (l_2) \log (l_3)
   \nonumber \\&-172 \zeta_3 \log
   (l_2)-\frac{412 \log ^4(l_2)}{3}-\frac{101}{3} \pi ^2 \log
   ^2(l_2)-24 \zeta_3 \log (l_3)-\frac{3 \log ^4(l_3)}{4}-\frac{3}{2}
   \pi ^2 \log ^2(l_3)-\frac{19 \pi ^4}{45}\,,
\\[4ex]
m_5^{(0)} &= 0 \,,\quad
m_5^{(1)} = 0 \,,\quad
m_5^{(2)} = 0\,, \nonumber \\
m_5^{(3)} &= -2 \text{Li}_3\left(-\frac{1}{l_4^2}\right)+\frac{4 \log
   ^3(l_4)}{3}+\frac{1}{3} \pi ^2 \log (l_4)+2 \zeta_3\,, \nonumber \\
m_5^{(4)} &= -8 \text{Li}_{2,2}\left(-1,-\frac{1}{l_4^2}\right)+4
   \text{Li}_4\left(-\frac{1}{l_1 l_4^2 l_5}\right)-4
   \text{Li}_4\left(\frac{l_1 l_5}{l_4^2}\right)-8 \log (l_1)
   \text{Li}_3\left(-\frac{1}{l_4^2}\right) 
   \nonumber \\&-16 \log (l_1)
   \text{Li}_3\left(\frac{1}{l_4^2}\right)+4 \log ^2(l_1) \log (l_4)
   \log (l_5)+4 \log (l_1) \log (l_4) \log ^2(l_5)
   \nonumber \\&-24 \log
   (l_1) \log ^2(l_4) \log (l_5)-12 \text{Li}_4\left(\frac{1}{l_1
   l_4}\right)-12 \text{Li}_4\left(\frac{l_4}{l_1}\right)+\frac{4}{3}
   \log ^3(l_1) \log (l_4)
   \nonumber \\&+\frac{64}{3} \log (l_1) \log ^3(l_4)-6
   \log ^2(l_1) \log ^2(l_4)-\frac{8}{3} \pi ^2 \log (l_1) \log
   (l_4)+\frac{2}{3} \log ^3(l_1) \log (l_5)
   \nonumber \\&+\frac{2}{3} \log
   (l_1) \log ^3(l_5)+\log ^2(l_1) \log ^2(l_5)+\frac{2}{3} \pi
   ^2 \log (l_1) \log (l_5)-4 \zeta_3 \log (l_1)-\frac{5}{6} \log
   ^4(l_1)
   \nonumber \\&+\frac{4}{3} \pi ^2 \log ^2(l_1)+8
   \text{Li}_2\left(-\frac{1}{l_4^2}\right) \text{Li}_2\left(-\frac{1}{l_4
   l_5}\right)+4 \text{Li}_2\left(-\frac{1}{l_4^2}\right) \log
   ^2(l_5)-8 \text{Li}_2\left(-\frac{1}{l_4^2}\right) \log (l_4) \log
   (l_5)
   \nonumber \\&-16 \text{Li}_3\left(-\frac{1}{l_4^2}\right) \log (l_5)-4
   \text{Li}_2\left(-\frac{1}{l_4^2}\right){}^2-\frac{4}{3} \pi ^2
   \text{Li}_2\left(-\frac{1}{l_4^2}\right)+12
   \text{Li}_4\left(-\frac{1}{l_4^2}\right)-4
   \text{Li}_4\left(\frac{1}{l_4^2}\right)
   \nonumber \\&-4
   \text{Li}_2\left(-\frac{1}{l_4^2}\right) \log ^2(l_4)+16
   \text{Li}_3\left(-\frac{1}{l_4^2}\right) \log (l_4)+\frac{2}{3} \pi ^2
   \text{Li}_2\left(-\frac{1}{l_4 l_5}\right)-16
   \text{Li}_4\left(-\frac{1}{l_4 l_5}\right)\nonumber \\&+16
   \text{Li}_4\left(\frac{l_5}{l_4}\right)+8 \log ^2(l_4)
   \text{Li}_2\left(-\frac{1}{l_4 l_5}\right)-\frac{4}{3} \log (l_4)
   \log ^3(l_5)+\frac{16}{3} \log ^3(l_4) \log (l_5)+4 \log
   ^2(l_4) \log ^2(l_5)
   \nonumber \\&+\frac{2}{3} \pi ^2 \log (l_4) \log
   (l_5)+12 \zeta_3 \log (l_4)-\frac{35 \log ^4(l_4)}{3}+\frac{1}{3}
   \pi ^2 \log ^2(l_4)-12 \zeta_3 \log (l_5)-\frac{\log
   ^4(l_5)}{2}
   \nonumber \\&-\frac{2}{3} \pi ^2 \log ^2(l_5)-\frac{\pi ^4}{72}\,,
\\[4ex]
m_6^{(0)} &= 0 \,,\quad
m_6^{(1)} = 0 \,, \nonumber \\
m_6^{(2)} &= 2 \text{Li}_2\left(-\frac{1}{l_4^2}\right)+2 \log ^2(l_4)+\frac{\pi ^2}{6}\,, 
\nonumber \\
m_6^{(3)} &= 4 \text{Li}_3\left(\frac{l_1 l_5}{l_4^2}\right)-8 \log (l_1)
   \text{Li}_2\left(-\frac{1}{l_4 l_5}\right)-8 \log (l_1) \log
   (l_4) \log (l_5)+4 \text{Li}_3\left(\frac{1}{l_1 l_4}\right)
   \nonumber \\&-6
   \log ^2(l_1) \log (l_4)+22 \log (l_1) \log ^2(l_4)-4 \log
   (l_1) \log ^2(l_5)-\frac{2}{3} \log ^3(l_1)-\pi ^2 \log
   (l_1)
   \nonumber \\&+10 \text{Li}_3\left(-\frac{1}{l_4^2}\right)+12
   \text{Li}_3\left(\frac{1}{l_4^2}\right)+8 \text{Li}_3\left(-\frac{1}{l_4
   l_5}\right)-4 \log (l_4) \log ^2(l_5)+20 \log ^2(l_4) \log
   (l_5)
   \nonumber \\&-\frac{70 \log ^3(l_4)}{3}+\frac{5}{3} \pi ^2 \log
   (l_4)-\frac{4 \log ^3(l_5)}{3}-\frac{4}{3} \pi ^2 \log (l_5)-8 \zeta
   (3)\,, \nonumber \\
m_6^{(4)} &= 4 \text{Li}_{2,2}\left(\frac{1}{l_1 l_4},\frac{l_1}{l_4}\right)-8
   \text{Li}_4\left(-\frac{1}{l_1 l_4^2 l_5}\right)-8
   \text{Li}_4\left(\frac{l_1 l_5}{l_4^2}\right)-2 \log ^2(l_1)
   \text{Li}_2\left(-\frac{1}{l_4^2}\right)
   \nonumber \\&+4 \log (l_1)
   \text{Li}_2\left(-\frac{1}{l_4^2}\right) \log (l_4)-16 \log (l_1)
   \text{Li}_3\left(\frac{1}{l_4^2}\right)+8 \log ^2(l_1)
   \text{Li}_2\left(-\frac{1}{l_4 l_5}\right)
   \nonumber \\&-32 \log (l_1)
   \text{Li}_3\left(-\frac{1}{l_4 l_5}\right)-16 \log (l_5)
   \text{Li}_3\left(\frac{1}{l_1 l_4}\right)+8 \log ^2(l_1) \log
   (l_4) \log (l_5)
   \nonumber \\&+8 \log (l_1) \log (l_4) \log ^2(l_5)-40
   \log (l_1) \log ^2(l_4) \log (l_5)+12
   \text{Li}_4\left(\frac{1}{l_1 l_4}\right)+16 \log (l_4)
   \text{Li}_3\left(\frac{1}{l_1 l_4}\right)
   \nonumber \\&+\frac{10}{3} \log ^3(l_1)
   \log (l_4)+46 \log (l_1) \log ^3(l_4)-21 \log ^2(l_1) \log
   ^2(l_4)-\frac{10}{3} \pi ^2 \log (l_1) \log (l_4)
   \nonumber \\&+\frac{4}{3} \log
   ^3(l_1) \log (l_5)+4 \log (l_1) \log ^3(l_5)+2 \log
   ^2(l_1) \log ^2(l_5)+\frac{8}{3} \pi ^2 \log (l_1) \log
   (l_5)+16 \zeta_3 \log (l_1)
   \nonumber \\&+\frac{\log ^4(l_1)}{6}+\pi ^2 \log
   ^2(l_1)-8 \text{Li}_3\left(-\frac{1}{l_4^2}\right) \log (l_5)-2
   \text{Li}_2\left(-\frac{1}{l_4^2}\right){}^2-40
   \text{Li}_4\left(-\frac{1}{l_4^2}\right)-28
   \text{Li}_4\left(\frac{1}{l_4^2}\right)
   \nonumber \\&-2
   \text{Li}_2\left(-\frac{1}{l_4^2}\right) \log ^2(l_4)+8
   \text{Li}_3\left(-\frac{1}{l_4^2}\right) \log (l_4)-\frac{2}{3} \pi ^2
   \text{Li}_2\left(-\frac{1}{l_4 l_5}\right)+16
   \text{Li}_4\left(-\frac{1}{l_4 l_5}\right)
   \nonumber \\&+16
   \text{Li}_4\left(\frac{l_5}{l_4}\right)+\frac{80}{3} \log ^3(l_4)
   \log (l_5)-4 \log ^2(l_4) \log ^2(l_5)-\frac{2}{3} \pi ^2 \log
   (l_4) \log (l_5)-8 \zeta_3 \log (l_4)
   \nonumber \\&-\frac{169 \log
   ^4(l_4)}{6}+\frac{2}{3} \pi ^2 \log ^2(l_4)+8 \zeta_3 \log
   (l_5)+\frac{\log ^4(l_5)}{3}+\frac{1}{3} \pi ^2 \log ^2(l_5)-\frac{7
   \pi ^4}{60}\,,
\\[4ex]
m_7^{(0)} &= 0 \,,\quad
m_7^{(1)} = 0 \,,\quad
m_7^{(2)} = 0\,, \nonumber \\
m_7^{(3)} &= 8 \log (l_1) \log ^2(l_2)-4 \text{Li}_3\left(\frac{1}{l_2^2}\right)-8
   \text{Li}_2\left(\frac{1}{l_2^2}\right) \log (l_2)-\frac{16}{3} \log
   ^3(l_2)+4 \zeta_3\,, \nonumber \\
m_7^{(4)} &= -24 \text{Li}_{2,2}\left(-1,-\frac{1}{l_2^2}\right)-24 \log (l_2)
   \text{Li}_3\left(-\frac{1}{l_1 l_2^2 l_3}\right)+16 \log (l_1)
   \text{Li}_2\left(\frac{1}{l_2^2}\right) \log (l_2)
   \nonumber \\&+48 \log (l_1)
   \text{Li}_3\left(\frac{1}{l_2 l_3}\right)+48 \log (l_1)
   \text{Li}_3\left(\frac{l_2}{l_3}\right)+12 \log ^2(l_1) \log
   (l_2) \log (l_3)
   \nonumber \\&+12 \log (l_1) \log (l_2) \log
   ^2(l_3)-24 \text{Li}_4\left(-\frac{1}{l_1 l_2}\right)+24
   \text{Li}_4\left(\frac{l_1}{l_2}\right)-36 \log (l_1) \log
   ^3(l_2)
   \nonumber \\&+2 \log ^2(l_1) \log ^2(l_2)-\frac{8}{3} \pi ^2 \log
   (l_1) \log (l_2)-16 \log (l_1) \log ^3(l_3)+8 \pi ^2 \log
   (l_1) \log (l_3)
   \nonumber \\&-84 \zeta_3 \log (l_1)-\log ^4(l_1)-2 \pi ^2
   \log ^2(l_1)+24 \text{Li}_2\left(\frac{1}{l_2^2}\right)
   \text{Li}_2\left(\frac{1}{l_2 l_3}\right)+12
   \text{Li}_2\left(\frac{1}{l_2^2}\right) \log ^2(l_3)
   \nonumber \\&+24
   \text{Li}_2\left(\frac{1}{l_2^2}\right) \log (l_2) \log (l_3)-4
   \text{Li}_2\left(\frac{1}{l_2^2}\right){}^2-\frac{2}{3} \pi ^2
   \text{Li}_2\left(\frac{1}{l_2^2}\right)+36
   \text{Li}_4\left(-\frac{1}{l_2^2}\right)-12
   \text{Li}_4\left(\frac{1}{l_2^2}\right)
   \nonumber \\&-20
   \text{Li}_2\left(\frac{1}{l_2^2}\right) \log ^2(l_2)-48
   \text{Li}_3\left(\frac{1}{l_2^2}\right) \log (l_2)+2 \pi ^2
   \text{Li}_2\left(\frac{1}{l_2 l_3}\right)+24 \log (l_2)
   \text{Li}_3\left(\frac{1}{l_2 l_3}\right)
   \nonumber \\&-24 \log (l_2)
   \text{Li}_3\left(\frac{l_2}{l_3}\right)+4 \log (l_2) \log
   ^3(l_3)-6 \pi ^2 \log (l_2) \log (l_3)+48 \zeta_3 \log
   (l_2)+19 \log ^4(l_2)
   \nonumber \\&+\frac{7}{3} \pi ^2 \log ^2(l_2)+\pi ^2 \log
   ^2(l_3)-\frac{14 \pi ^4}{45}\,,
\\[4ex]
m_8^{(0)} &= 0 \,,\quad
m_8^{(1)} = 0 \,,\quad
m_8^{(2)} = 0\,, \quad
m_8^{(3)} = 0\,, \nonumber \\
m_8^{(4)} &= -8 \text{Li}_4\left(-\frac{1}{l_1 l_2}\right)+8
   \text{Li}_4\left(\frac{l_1}{l_2}\right)+16 \log (l_2)
   \text{Li}_3\left(\frac{l_1}{l_2}\right)-\frac{4}{3} \log ^3(l_1)
   \log (l_2)+\frac{44}{3} \log (l_1) \log ^3(l_2)
   \nonumber \\&-2 \log ^2(l_1)
   \log ^2(l_2)-\frac{4}{3} \pi ^2 \log (l_1) \log (l_2)-\frac{1}{3}
   \log ^4(l_1)-\frac{2}{3} \pi ^2 \log ^2(l_1)+8
   \text{Li}_3\left(\frac{1}{l_2^2}\right) \log (l_2)
   \nonumber \\&-8 \zeta_3 \log
   (l_2)-\frac{29 \log ^4(l_2)}{3}+2 \pi ^2 \log ^2(l_2)-\frac{7 \pi
   ^4}{45}\,,
\\[4ex]
m_9^{(0)} &= 0 \,,\quad
m_9^{(1)} = 0 \,,\quad
m_9^{(2)} = 0\,, \quad
m_9^{(3)} = 0\,, \nonumber \\
m_9^{(4)} &= 4 \text{Li}_4\left(-\frac{1}{l_1 l_2}\right)-4
   \text{Li}_4\left(\frac{l_1}{l_2}\right)-8 \log (l_2)
   \text{Li}_3\left(\frac{l_1}{l_2}\right)+\frac{2}{3} \log ^3(l_1)
   \log (l_2)-\frac{22}{3} \log (l_1) \log ^3(l_2)
   \nonumber \\&+\log ^2(l_1)
   \log ^2(l_2)+\frac{2}{3} \pi ^2 \log (l_1) \log (l_2)-4 \log
   (l_1) \text{Li}_3\left(-\frac{1}{l_4^2}\right)+\frac{8}{3} \log
   (l_1) \log ^3(l_4)
   \nonumber \\&+\frac{2}{3} \pi ^2 \log (l_1) \log (l_4)+4
   \zeta_3 \log (l_1)+\frac{\log ^4(l_1)}{6}+\frac{1}{3} \pi ^2 \log
   ^2(l_1)-4 \text{Li}_3\left(\frac{1}{l_2^2}\right) \log (l_2)+4 \zeta
   (3) \log (l_2)
   \nonumber \\&+\frac{29 \log ^4(l_2)}{6}-\pi ^2 \log ^2(l_2)-2
   \text{Li}_2\left(-\frac{1}{l_4^2}\right){}^2-\frac{1}{3} \pi ^2
   \text{Li}_2\left(-\frac{1}{l_4^2}\right)-4
   \text{Li}_2\left(-\frac{1}{l_4^2}\right) \log ^2(l_4)
   \nonumber \\&-2 \log
   ^4(l_4)-\frac{1}{3} \pi ^2 \log ^2(l_4)+\frac{23 \pi ^4}{360}\,,
\end{align}